\documentclass{pasj01}
\usepackage{color}

\Received{}
\Accepted{}
 
\begin{document} 

\title{A light curve model of V2491 Cyg:
Classical nova outburst on a cool and massive white dwarf
%
}

\author{Mariko \textsc{Kato}\altaffilmark{1}}
\altaffiltext{1}{Department of Astronomy, Keio University, Hiyoshi, Yokohama
  223-8521, Japan}
\email{mariko.kato@hc.st.keio.ac.jp}

\author{Hideyuki \textsc{Saio},\altaffilmark{2}}
\altaffiltext{2}{Astronomical Institute, Graduate School of Science,
    Tohoku University, Sendai, 980-8578, Japan}

\author{Izumi \textsc{Hachisu}\altaffilmark{3}}
\altaffiltext{3}{Department of Earth Science and Astronomy, College of Arts and
Sciences, The University of Tokyo, 3-8-1 Komaba, Meguro-ku, Tokyo 153-8902, Japan}

\maketitle

\begin{abstract}
The classical nova V2491~Cyg was once suggested to be a recurrent nova. 
We have broadly reproduced the light curve of V2491 Cyg
by a nova outburst model on a cold $1.36~M_\odot$ white dwarf (WD), which  
strongly suggests that V2491~Cyg is a classical nova outbursting
on a cold very massive WD rather than a recurrent nova outbursting
on a warmer WD like the recurrent nova RS~Oph.
In a long-term evolution of a cataclysmic binary, an accreting 
WD has been settled down to a thermal equilibrium state with the balance of
gravitational energy release and neutrino loss.  The central temperature
of the WD is uniquely determined by the energy balance. 
The WD is hot (cold) for a high (low) mass-accretion rate. 
We present the central temperatures, ignition masses, ignition radii, and 
recurrence periods for various WD masses and mass-accretion rates. 
In a classical nova, which corresponds to a low mass-accretion rate, 
the WD is cool and strongly degenerated and the ignition mass is large, 
which result in a strong nova outburst. 
In a recurrent nova, the WD is relatively warmer because of a high mass
accretion rate and the outburst is relatively weaker. 
The gravitational energy release substantially contributes to the 
luminosity during the recurrent nova outbursts.  We compare physical
properties between classical novae and recurrent novae and discuss
the essential differences between them.
\end{abstract}


\section{Introduction}

A nova occurs on an accreting white dwarf (WD) in a cataclysmic binary system. 
When the accreted H-rich matter reaches a critical value (ignition mass) 
hydrogen thermonuclear runaway sets in, which triggers a nova outburst. 
The envelope expands to a giant size and strong wind mass-loss occurs. 
The optical brightness quickly rises to maximum followed by a gradual 
decrease. 
After the wind stops, the WD enters the supersoft X-ray source (SSS) phase
that continues until the hydrogen burning extinguishes. 

\citet{kat83} proposed an idea that the nova decay phase can be followed 
by a sequence of steady-state solutions.  In nova outbursts,
winds are accelerated deep inside the photosphere.  
Based on the optically thick wind theory together with the OPAL opacity
\citep{igl96}, \citet{kat94h} systematically calculated nova outburst 
evolutions for various WD masses and chemical compositions.
Using their results, many nova light curves have been calculated,
which are in good agreement with observed light curves for a number 
of classical novae 
\citep{hac06k, hac15k, hac16a, hac18a, hac18b, hac19b, hac21k}. 
In these models, optical/IR light curves are followed
by the free-free emission light curves based on the wind mass-loss rates
obtained with the optically thick wind theory. 
After the wind stops, the SSS phase can be followed by the sequence 
of static solutions.  Good agreements of theoretical
with observed light curves indicate that the steady-state 
assumption in the optically thick winds can be well applied
to the decay phase of a classical nova. 

The timescale of a nova outburst evolution depends strongly
on the WD mass but weakly on the chemical composition of the envelope.
Thus, the decline rate in the optical light curve and duration of
a SSS phase give a rough estimate of the WD mass. 
In general a fast (slow) nova corresponds to a massive (less massive) WD
(e.g., \citet{hac20}).   
Both the optical and SSS phases have been theoretically calculated 
for various WD masses and envelope chemical compositions 
\citep{kat94h, kat99, sal05, hac06k, wol13, kat13hh, hac15k,
hac18b, hac19b, kat20sh, hac21k}. 
The overall light curves of IR/optical and SSS phases are well reproduced 
with the optically thick wind theory for a number of classical novae 
\citep{kat94h, kat99, hac06k, kat13hh, hac15k, hac18b, hac19b, hac21k}. 

V2491 Cyg outbursted in 2008 \citep{nak08}.   
Its optical light curve shows a rapid decline followed
by a plateau-like phase, resembling those of recurrent novae. 
The short SSS phase suggests a very massive WD. 
\citet{nes11} pointed out the strong similarity of X-ray spectrum 
in V2491 Cyg and the recurrent nova RS Oph. 
\citet{pag10} and \citet{pag14} discussed that V2491 Cyg is possibly
a recurrent nova (see, also, \citet{tom08}). 
\citet{hac19a} compared the light curves of V2491 Cyg and RS Oph,
both of which show a similar X-ray emergence time (see their figure 16).  
These similarities suggest that both RS Oph and V2491 Cyg host a massive WD ($\sim 1.35 ~M_\odot$: see the next section). 
On the other hand, it was suggested that V2491 Cyg should be 
a classical nova, 
because (1) a heavy element enrichment is detected in its ejecta 
\citep{mun11}, which is a characteristic property of classical novae,
and also because (2) V2491 Cyg seems to be an intermediate polar (IP) 
\citep{tak09, tak11, zem15, sun20} whereas   
no recurrent novae have been identified as an IP.

\citet{kat20} found a statistical difference between classical novae 
and recurrent novae in the $t_{\rm off}- t_{\rm SSS}$ diagram 
in which recurrent novae have a longer SSS duration 
than classical novae do. 
Here $t_{\rm off}$ is the X-ray turnoff time, which indicates the total 
duration of the outburst, and $t_{\rm SSS}$ is the SSS duration.  
They showed that the recurrent nova RS Oph is on the trend of 
recurrent nova relation, and V2491 Cyg is on the classical nova relation. 
These two novae show a similar X-ray turn-on time but different SSS durations. 

In the present paper, we clarify the reason for the different 
SSS durations between the classical novae and recurrent novae 
both having similar WD masses. The main difference between them is the 
mass accretion rate. Thus, we have calculated the interior structure 
of accreting WDs for various mass-accretion rates.

This paper is organized as follows. 
Section \ref{sec_rsophv2491} lists the resemblance and difference 
between the observational light curves of RS Oph and V2491 Cyg. 
Section \ref{sec_interior} presents our calculation for 
interior structures of mass accreting WDs for various mass-accretion 
rates.  Based on the results, we present theoretical light curves of V2491 Cyg 
as a classical nova on a very cold WD in section \ref{section_steadystate}. 
The light curve is well explained by the optically thick wind theory 
and additional emission from massive ejecta.  
In section \ref{sec_evolution},  we study how and why the recurrent nova
outbursts are different from the steady-state solutions. 
Discussion and conclusions follow in sections \ref{sec_discussion} and 
\ref{section_conclusion}, respectively.

\section{Observational Properties of V2491~Cyg and RS Oph}
\label{sec_rsophv2491}

\begin{figure*}
  \includegraphics[width=12cm]{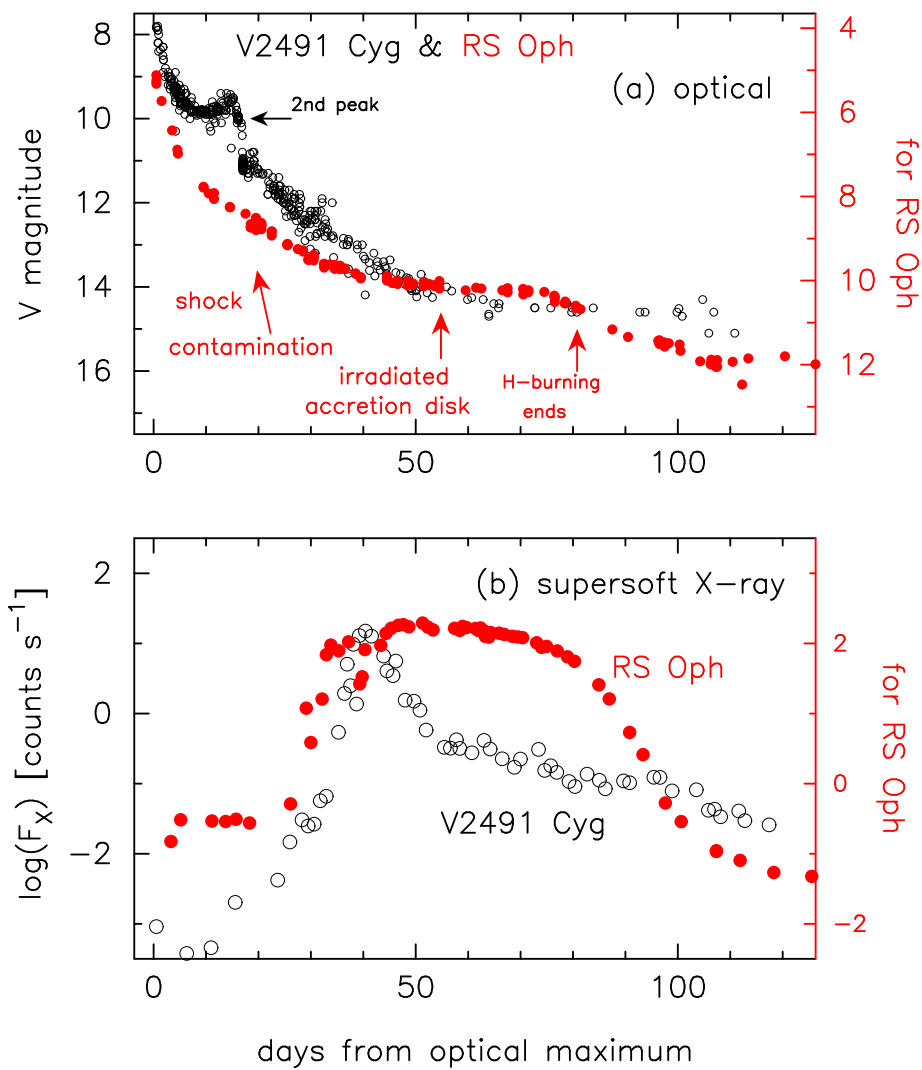}
\caption{
(a) The $V$ light curves and (b) the supersoft X-ray count rates
both for V2491~Cyg (black symbols) and RS~Oph (red symbols). 
The $V$ light curve of V2491 Cyg is shifted upward by 3.85 mag to 
match the absolute magnitudes of RS Oph. 
Also the X-ray count rates of V2491 Cyg are shifted upward by 1.0 dex
to match the peak count rates.
See the main text for details. 
}\label{v2491cygrsoph}
\end{figure*}

Figure \ref{v2491cygrsoph} shows the $V$-band and supersoft X-ray
light curves of the classical nova V2491~Cyg (black symbols) as well as
those of the recurrent nova RS Oph (red symbols) in the 2006 outburst.
For RS Oph, the optical data are taken from \citet{hac06c} and X-ray data 
are from \citet{hac07kl}.  The optical decline of RS Oph is smooth,
followed by a plateau phase.  Such an optical light curve is modeled
by \citet{hac06c} in which the early decline is mainly described 
by the free-free emission from optically thin ejecta just outside 
the photosphere of the WD envelope, whereas the plateau is caused
mainly by the photospheric emission from an irradiated optically thick 
accretion disk.  Their model consists of a $1.35~M_\odot$ WD, 
a large disk with a radius of $\sim 40~R_\odot$ and a red giant companion.  
The disk is so large because RS Oph is a symbiotic recurrent nova 
of orbital period 453.6 days \citep{bra09} and the separation is 
as large as $\sim 300~R_\odot$ that permits such a large accretion disk. 
In classical novae, on the other hand, the orbital period is typically
a few to several hours, 
and the disk is too small to contributes much to the $V$ light curve. 
\citet{hac06c} identified the epoch when the hydrogen burning turned off 
as indicated by a red arrow in figure \ref{v2491cygrsoph}(a). 
Then, the optical flux decayed faster as the X-ray count rate declines.  

We should note that \citet{iijima09} differently interpreted  
that the decay of supersoft X-ray and optical flux in 80-90 days 
to be the shock breakout from the cool red-giant wind based 
on his optical spectral analysis of the RS~Oph 2006 outburst.

The reddening and distance to RS Oph were obtained to be $E(B-V)=0.65$ mag, 
$(m-M)_V=12.8$ mag and $d=1.4$ kpc by \citet{hac18b} and \citet{hac21k}. 
Recently, \citet{hac21k} examined multiwavelength light curves of RS Oph,
and clarified that the $V$ band light curve of RS Oph is contaminated with 
shock radiation and thus decays somewhat slower than the pure free-free
emission light curve (see their figure 20).

V2491 Cyg, outbursted in 2008, is a fast nova with a moderately bright 
second peak in its optical light curve. 
The light curves are plotted in figure \ref{v2491cygrsoph} (black symbols), 
the data of which are the same as those in \citet{hac09}. 
This object is strongly suggested to be an intermediate polar (IP)
\citep{tak09,tak11,zem15}. 
For the mechanism of the second peak \citet{hac09} proposed 
an idea that stored magnetic energy is released 
when the magnetic energy becomes larger than the rotation energy of 
the expanding envelope, which should occur in the middle of the optical decline.   
\citet{hac16a} examined V2491 Cyg in the $UBV$ color evolution 
and obtained $E(B-V)=0.23$, $(m-M)_V=16.5$. $d=14$ kpc.
These values are revised to be 
$E(B-V)=0.4\pm 0.05$, $(m-M)_V=16.65\pm0.2$, $d=12.1 \pm 2$ kpc
based on the time-stretched color-magnitude diagram method \citep{hac21k}.

V2491 Cyg was well observed with X-ray satellites.  
The X-ray light curve and spectra are reported in detail 
by \citet{pag10} and \citet{nes11}. 
The X-ray light curve shows a rather shorter ($\sim 10$ days) bright
peak than that of RS~Oph, followed by a decline phase. 
The decline rate (slope) suddenly becomes slower on Day $\sim 53$.
This change has not been theoretically explained 
while many novae show a quick decay in the X-ray count rate, e.g., 
in V1974 Cyg, V382 Vel, and V597 Pup \citep{hac10k}. 
X-ray spectra were taken on Day $\sim 40$ and Day $\sim 50$
with {\it XMM-Newton}. 
\citet{nes11} pointed out the remarkable similarity of the X-ray spectra 
between V2491 Cyg on Day 40.18 and RS Oph on Day 39.7. 
Also the blackbody temperature $6.3 \times 10^5$ K 
of V2491 Cyg on Day 40.18 \citep{nes11}
is similar to $(6.3-8.3) \times 10^5$ K of RS Oph on Day 39.7 \citep{nes07}. 
Such similarities suggest that the both novae have similar mass WDs. 
The mass estimate in the theoretical light curve fittings 
indicates $\sim 1.35~M_\odot$ for the both novae.  See figure 48 of
\citet{hac19a} for V2491 Cyg, and figure 3 of \citet{hac06c} for RS Oph.  

Figure \ref{v2491cygrsoph}a compares the $V$ light curves of the two novae.   
The $V$ light curve of V2491 Cyg are shifted upward by 3.85 mag to match the 
absolute magnitudes of RS Oph. Here 3.85 mag corresponds to the difference 
of the two distance moduli, that is, $(m-M)_{\rm V}=16.65$ for V2491 Cyg 
\citep{hac21k} and $(m-M)_V=12.8$ for RS Oph \citep{hac18b}. 
V2491 Cyg shows a prominent second peak which has not been observed
in recurrent novae.  Also the bright optical peak of V2491 Cyg, 
$M_V=-9.2$ mag, is much brighter than $M_V=-8.0$ mag in RS Oph or 
other recurrent novae \citep{hac18b}. 

\citet{hac20} theoretically related the peak brightness ($M_{V, \rm max}$)
and decline time ($t_3$) by 3 mag from the peak to the ignition mass
and mass accretion rate (or recurrence period).  If we assume that
$M_{V,\rm max}=-9.2$ and $t_3= t_2/0.54= 4/0.54= 7.4$~d for V2491~Cyg, 
their figure 6 indicates that the WD mass is $M_{\rm WD}= 1.32~M_\odot$
and the mass accretion rate to the WD is 
$\dot M_{\rm acc}= 3\times 10^{-11} ~M_\odot$~yr$^{-1}$ 
or the recurrence period is $P_{\rm rec}= 1.5 \times 10^5$~yr
for solar abundance.
These properties indicate that V2491 Cyg is a classical nova rather 
than a recurrent nova.

After Day 60, the optical light curve of V2491 Cyg becomes flat, 
resembling to a plateau phase in recurrent novae.
The absolute $V$ magnitude of the flat phase in V2491~Cyg is
slightly fainter than that in RS~Oph (see figure \ref{v2491cygrsoph}a).
In recurrent novae, the plateau phase is explained by the contribution 
of an irradiated accretion disk. The plateau phase is brighter for a 
larger accretion disk.  The brightness of plateau is about
$M_V= m_V - (m-M)_V \approx 14.5 - 16.65= -2.15$ in V2491~Cyg,
$M_V= m_V - (m-M)_V \approx 14.5 - 16.85= -2.35$ in U~Sco \citep{hac21k}, and
$M_V= m_V - (m-M)_V \approx 10.0 - 12.8= -2.8$ in RS~Oph.
This trend simply means that the size of accretion disk is smallest 
in V2491~Cyg and largest in RS~Oph, and the size in U~Sco is between them. 
The orbital period of U~Sco is $P_{\rm orb}=1.23$~d, so we expect
that the orbital period of V2491~Cyg is probably shorter than a day.

V2491 Cyg is strongly suggested to be a member of intermediate polars
(IPs) \citep{zem15}, of which the typical orbital period ranges
from a few to several hours.\footnote{The Catalog of IPs and
IP Candidates by Right Ascension in 
https://asd.gsfc.nasa.gov/Koji.Mukai/iphome/catalog/alpha.html by 
K. Mukai (2014). Among 74 IPs and IP candidates listed
with its orbital period, only two have the orbital period larger
than 10 hr: 47.9 hr (GK Per) and 15.4 hr (V2731 Oph).}  
We suppose that the orbital period is a few to several hours
and therefore the orbital separation of V2491 Cyg is probably
as small as $\lesssim 1.5~R_\odot$.   
A small irradiated accretion disk hardly contributes to the visual band. 
For example, the recurrent nova U Sco ($P_{\rm orb}=1.23$~d) has  
an irradiated disk of $\sim 2~R_\odot$ \citep{hkkm00}, of which 
the contribution to the total brightness is much fainter than 
that in RS Oph (see e.g., figure 124 in \citet{hac19b}).

\section{Accreting White Dwarfs: Hot or Cool ?}
\label{sec_interior}

In many nova calculations, the central temperature of the WD and 
the mass accretion rate are often treated as independent parameters
(e.g. \citet{pri95,yar05,sta12,wol13,den13,che19}). 
The resultant flash properties depend largely on the WD temperature 
even if the mass-accretion rate and WD mass are the same. 
In a long term binary evolution, accreting WDs 
undergo a number of shell flashes, and 
their interiors adjust to their mass accretion 
(for more details, see section \ref{discussion_Tc}). 
To clarify the difference between V2491 Cyg and RS Oph, 
we need to obtain their WD models consistent with 
their mass-accretion rates, because their WD masses are probably
the same or very similar but their mass accretion rates are
very different.

We calculate hydrogen shell flash models for each set of the WD mass 
and mass accretion rate.  To prepare an initial WD interior model,
we obtain a thermal equilibrium model of each WD by balancing 
the heating by mass-accretion with the cooling by neutrino loss. 
Our calculation method and a part of the results have been already
published in \citet{kat14shn}, \citet{hac16sk}, and \citet{hac20}. 
We obtain the hydrogen-rich envelope mass at ignition,
$M_{\rm ig}$, central temperature of the WD, $T_{\rm c}$, 
and radius of the bottom of hydrogen burning zone, $R_{\rm b}$, 
at the onset of the shell flash (more exactly, at the epoch when the 
nuclear burning rate increases to $L_{\rm nuc}=10^6~L_\odot$).  
The accreted matter has the solar composition of $X=0.7$, $Y=0.28$,
and $Z=0.02$ by mass fraction.  Our results are summarized
in figures \ref{dmdtmig}, \ref{Twd}, and \ref{WDradius}.

\subsection{Ignition Mass} \label{sec_mig}

\begin{figure*}
 \begin{center}
  \includegraphics[width=12cm]{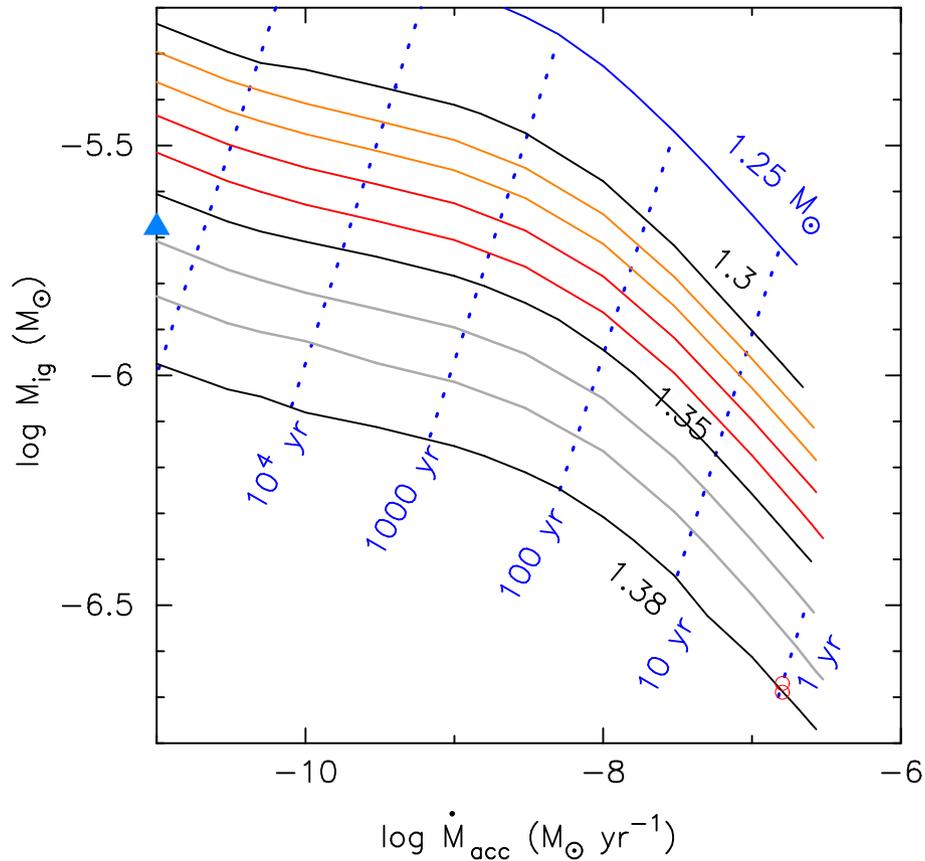}
 \end{center}
\caption{
The ignition mass vs. mass accretion rate for the WD mass of, from upper to 
lower,  1.25 (blue line), 1.3 (black), 1.31 and 1.32 (orange), 1.33 and
1.34 (red), 1.35 (black), 1.36 and 1.37 (gray), and 1.38 (black) $M_\odot$. 
The two red open cycles indicate the ignition mass of evolution model 
of 1.38 $M_\odot$ with $\dot M_{\rm acc}=1.6 \times 10^{-7}~M_\odot$~yr$^{-1}$ 
at the 10th flash (lower) \citep{kat17sha} and 1543th flash (upper)
\citep{kat17shb}.
The filled blue triangle indicates the envelope mass at the optical peak of 
the light curve model of V2491 Cyg (section \ref{sec_v2491cyg}). 
The dotted blue line connects loci at which the recurrence 
period $P_{\rm rec}$ is constant. 
}\label{dmdtmig}
\end{figure*}

Figure \ref{dmdtmig} shows the ignition mass vs. mass accretion rate for 
various WD masses.  The ignition mass is smaller for massive WDs
and for high mass-accretion rates. 
The gravitational energy release rate during a mass accretion phase
is larger in more massive WDs
and for larger mass-accretion rates, which keeps the envelope warmer 
to reduce the ignition mass (see \citet{kat14shn} for details).
The slope of the ignition mass vs. $\dot M_{\rm acc}$ line along the same
WD mass becomes steeper for 
$\dot M_{\rm acc} \gtrsim 10^{-8}~M_\odot$~yr$^{-1}$,  
because the envelope is not
degenerated in such high mass-accretion rates as shown later,
and the envelope is easily heated enough to ignite hydrogen. 
Note that the shell flash can occur even in a non-degenerate envelope
(see Discussion in section \ref{shellflash}).
On the other hand, in low mass-accretion rates, 
the accretion heating is not effective because the envelope is strongly
degenerated and efficient conductive energy transport 
carries heat away in a long interflash period. 
Thus, much mass need to be accumulated until it satisfies the ignition 
condition with help of blanket effect. 

The dotted blue lines in figure \ref{dmdtmig} indicate
the recurrence period calculated from 
$(M_{\rm ig}-M_{\rm 0}) / \dot M_{\rm acc}$, 
where $M_{\rm 0}$ is leftover of the hydrogen-rich envelope 
remained unburnt at the end of the previous outburst. 
This $M_{\rm 0}$ is calculated from the sequence of static solutions
at the epoch when the hydrogen burning turns off \citep{kat99}. 
The WD radius is assumed to be $R_{\rm b}$ in figure \ref{WDradius}. 
In low mass accretion rates, the recurrence period is 
approximately equal to $M_{\rm ig}/\dot M_{\rm acc}$ 
because $M_{\rm 0}$ is much smaller than $M_{\rm ig}$. 
The recurrent novae correspond to high mass-accretion rates 
$\dot M_{\rm acc}\sim 10^{-7}~M_\odot$~yr$^{-1}$,  
whereas the classical novae to 
$\dot M_{\rm acc}=10^{-11}$ to $10^{-8} M_\odot$~yr$^{-1}$.

The two red circles in figure \ref{dmdtmig} denote the ignition masses
taken from our successive hydrogen flash models of a $1.38~M_\odot$ WD
with $\dot M_{\rm acc}=1.6 \times 10^{-7}~M_\odot$~yr$^{-1}$. 
The lower circle is for the 10th flash \citep{kat17sha}, in which 
$M_{\rm ig}=2.0 \times 10^{-7}~M_\odot$, and $P_{\rm rec}= 0.95$~yr.  
The upper circle is for the last hydrogen flash  (the 1543th flash)
before the helium shell flash occurs,
$M_{\rm ig}=2.1 \times 10^{-7}~M_\odot$ and $P_{\rm rec}= 0.91$~yr. 
Both of them locate very close to the line of our 1.38 $M_\odot$ WD,  
because the flash properties are hardly changed during the 
evolution.

Figure \ref{dmdtmig} tells us a difference between the classical novae
and recurrent novae with the same WD mass.  
For example, in a $1.35~M_\odot$ WD with 
$\dot M_{\rm acc} = 5 \times 10^{-8}~M_\odot$~yr$^{-1}$ 
($P_{\rm rec} = 12$~yr), the ignition mass of a recurrent nova 
is $M_{\rm ig}=7.05 \times 10^{-7}~M_\odot$.  
On the other hand, a classical nova of $1.0\times 10^{-9}~M_\odot$~yr$^{-1}$ 
($P_{\rm rec} = 1650$~yr), 
the ignition mass is $M_{\rm ig}=1.65 \times 10^{-6}~M_\odot$, 2.3 
times larger than the recurrent nova. 
In an extremely low mass-accretion rate,  
$\dot M_{\rm acc}=1.0\times 10^{-11}~M_\odot$~yr$^{-1}$ 
($P_{\rm rec} = 2.48 \times 10^5$~yr), 
the ignition mass $M_{\rm ig}=2.48 \times 10^{-6}~M_\odot$ is 4.5 times
larger than that of the recurrent nova. 
These larger ignition masses guarantee massive ejecta.
Because the optical $V$ peak is brighter for massive ejecta 
(\citet{hac20} and section \ref{sec_v2491cyg} in the present paper), 
a classical nova is much brighter than a recurrent nova in its 
optical peak even if the WD masses are the same.

\subsection{White Dwarf Temperature} \label{section_Twd}

\begin{figure*}
 \begin{center}
  \includegraphics[width=12cm]{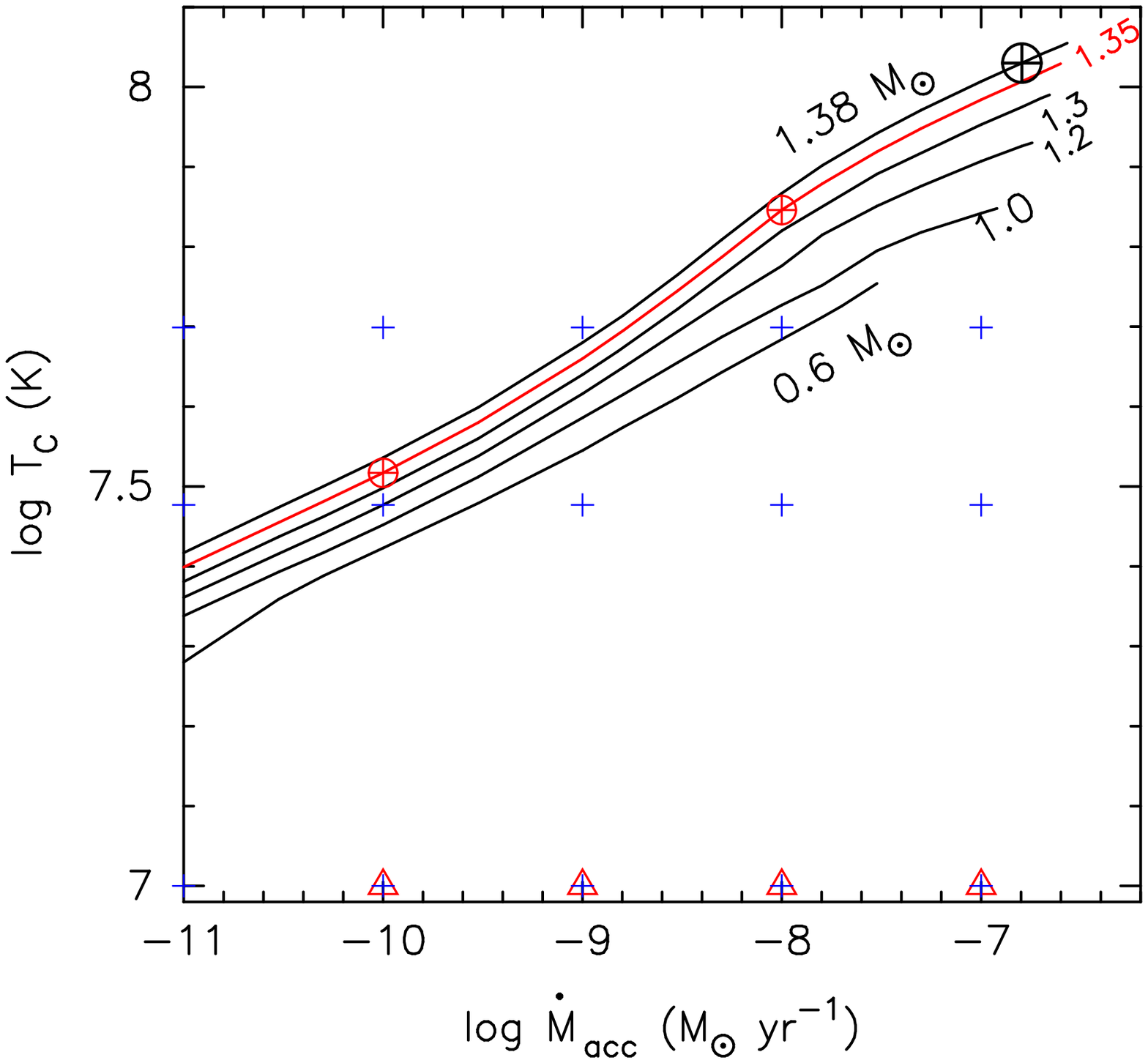}
 \end{center}
\caption{
The central temperature $T_{\rm c}$ vs. 
mass accretion rate $\dot M_{\rm acc}$ of the models in 
figure \ref{dmdtmig}. From lower to upper, 
0.6, 1.0, 1.2, and $1.3~M_\odot$ (black line),
1.35 $M_\odot$ (red line), and $1.38~M_\odot$ (black line) WDs.  
The black encircled plus symbol denotes the central temperature of 
a $1.38~M_\odot$ WD with 
$\dot M_{\rm acc}=1.6 \times 10^{-7}~M_\odot$~yr$^{-1}$ 
after it had experienced 10 and 1543 hydrogen shell flashes 
(two black symbols are overlapped  because the central temperature 
hardly changed). 
The two red encircled plus symbols indicate the 1.35 $M_\odot$ models 
in Figure \ref{m135interior} with 
$\dot M_{\rm acc}=1 \times 10^{-10}$ and $1\times 10^{-8}~M_\odot$~yr$^{-1}$. 
The blue cross indicates the assumed core temperature, 
1, 3, and 5 $\times 10^7$~K in \citet{pri95} and \citet{yar05}. 
The open red triangles are in \citet{che19}. 
}\label{Twd}
\end{figure*}


Figure \ref{Twd} summarizes the temperature at the center of the WD, 
$T_{\rm c}$, against various WD masses and mass-accretion rates. 
The central temperature is determined from a thermal balance between the
gravitational energy release owing to mass-accretion and neutrino loss.  
The core temperature is higher for a larger mass-accretion rate.  
The slope of $\Delta \log T_{\rm c}/\Delta \log \dot M_{\rm acc}$ is 
moderate for $\dot M_{\rm acc} > 10^{-8}~M_\odot$~yr$^{-1}$ 
because neutrino loss increases with the temperature. 

The black encircled plus symbol denotes the central temperature of 
the $1.38~M_\odot$ WD model with 
$\dot M_{\rm acc} =1.6 \times 10^{-7}~M_\odot$~yr$^{-1}$
after it experienced successive 10 and 1543 hydrogen flashes
in our long-term calculation \citep{kat17sha,kat17shb}. 
Starting with the steady-state (thermal equilibrium) model, 
the central temperature did not change during the series of shell flashes. 
Thus, the central temperatures at the two epochs are identical.     
This point lies on our theoretical $1.38~M_\odot$ WD line in figure \ref{Twd}. 
Note that the core temperature hardly changes during one cycle of
shell flash even when the temperature at the nuclear burning region undergoes
a cyclic change between $\log T$ (K)=7.68 and 8.25.

The blue crosses and red triangles are the WD temperatures 
adopted in shell flash calculations \citep{pri95,yar05,che19}, which 
will be discussed in section \ref{discussion_Tc}.


\subsection{Radii of Accreting White Dwarfs}
\label{sec_WDradius}

\begin{figure*}
  \includegraphics[width=10cm]{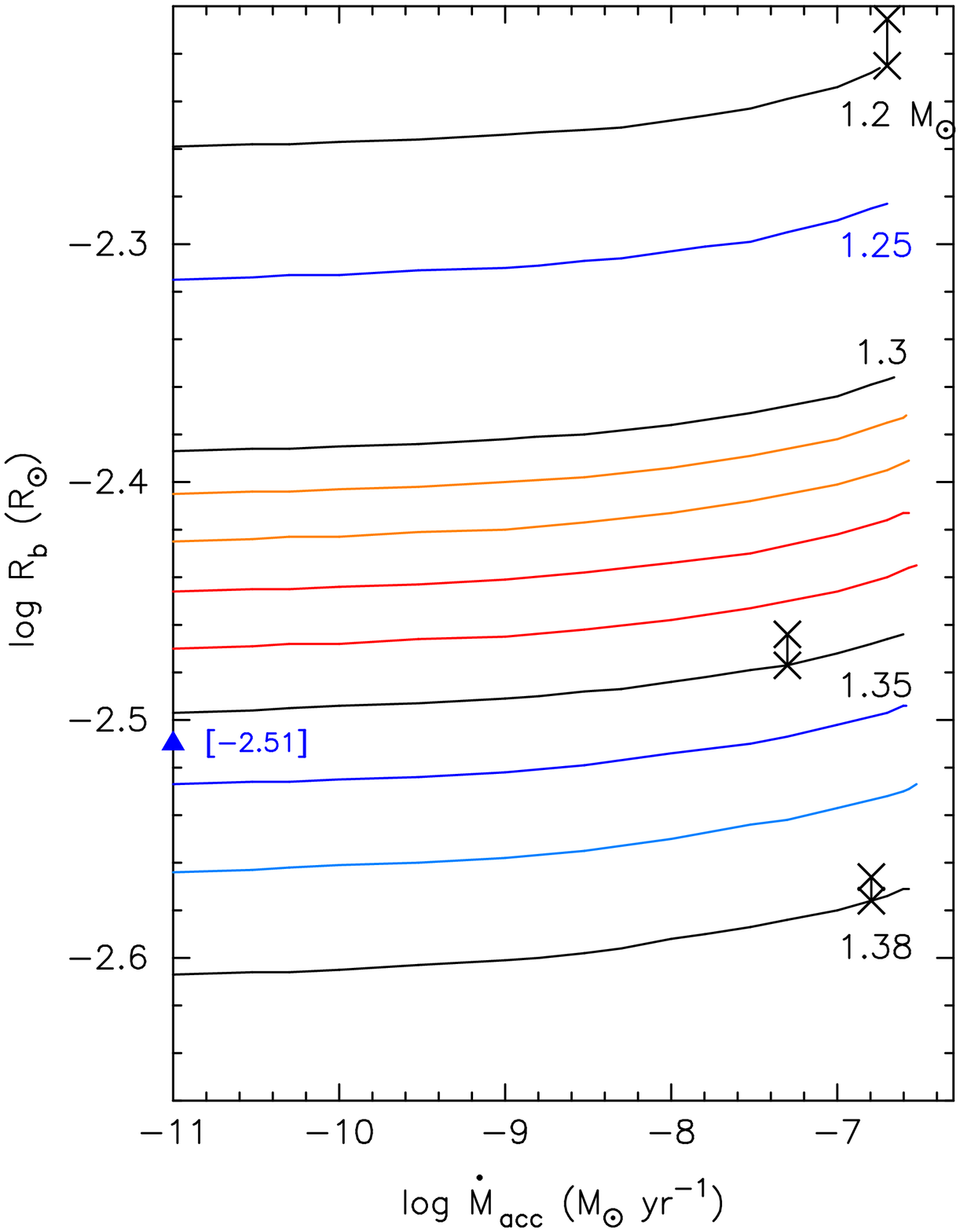}
\caption{
The radius $R_{\rm b}$, where the ignition occurs at the bottom of the 
H-rich envelope, is shown against the mass accretion rate. 
From upper to lower, the WD mass is 1.2 (black solid line), 1.25 (blue),  
1.3 (black), 1.31 and 1.32 (orange), 1.33 and 1.34 (red), 1.35 (black), 
1.36 (blue), 1.37 (sky blue), and 1.38 $M_\odot$ (black). 
The three pairs of X symbols connected with a line are taken from
the models in figure \ref{LnLg3mass}. 
The lower/upper symbol denotes the radius before/after the shell flash.
See section \ref{sec_3RN} for more details. 
}\label{WDradius}
\end{figure*}

Figure \ref{WDradius} shows the radius of the bottom of the H-rich envelope 
at the beginning of a hydrogen shell flash. 
In general, the WD radius is smaller in a more massive WD, and rapidly
decreases as the WD mass approaches the Chandrasekhar mass limit. 
The radius in figure \ref{WDradius} follows this tendency 
and, in addition, it shows a dependence on the mass-accretion rate. 
The radius is slightly larger for a larger mass-accretion rate  
because the WD is warmer by a larger gravitational energy release rate. 
Comparing a classical nova with a recurrent nova having the same WD mass,
we expect slightly stronger shell flashes in low mass-accretion rates 
because the surface gravity is stronger in a smaller radius.

\begin{figure*}
  \includegraphics[width=12cm]{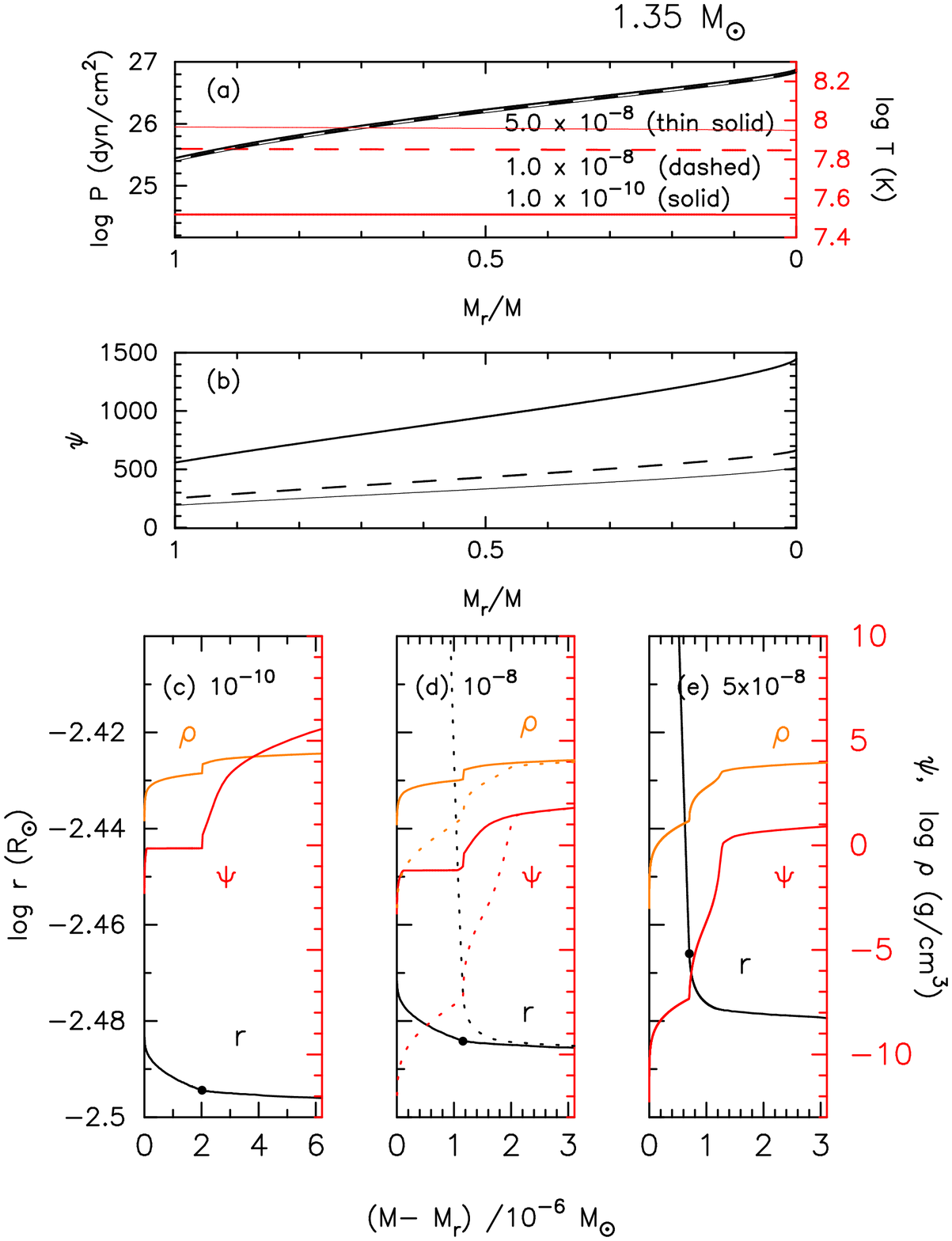}
\caption{
Various interior structures of mass-accreting 1.35 $M_\odot$ WDs.
(a) The distributions of the pressure $P$ (black lines) and 
temperature $T$ (red lines) against the mass coordinate $M_r/M$, where
$M_r$ is the mass within the radius $r$ and $M$ is the total mass
of the WD.  (b) The degeneracy $\psi$ from the WD center (the right edge)
to the surface (the left edge).  In panels (a) and (b), the accretion rate is 
$\dot M = 1 \times 10^{-10}~M_\odot$~yr$^{-1}$ (thick solid line), 
$\dot M = 1 \times 10^{-8}~M_\odot$~yr$^{-1}$ (dashed line), and
$\dot M = 5 \times 10^{-8}~M_\odot$~yr$^{-1}$ (thin solid line). 
(c)(d)(e) The bottom three panels show the degeneracy $\psi$ (red),
radius $r$ (black), and density $\rho$ (orange)
against $(M-M_r)/10^{-6}M_\odot$, i.e.,
in the very surface layer for the three different mass accretion rates,
$1 \times 10^{-10}$, $1 \times 10^{-8}$, and
$5 \times 10^{-8} ~M_\odot$~yr$^{-1}$, respectively.  
The left edge of each figure corresponds to the surface of 
accreted H-rich envelope, the bottom of which is indicated by
the black dot on the $r$ distribution. 
See the main text for other details.
}\label{m135interior}
\end{figure*}

\subsection{Interior Structures of Accreting 1.35 $M_\odot$ WDs}
\label{sec_wd}

Figure \ref{m135interior} shows the interior structures 
at the onsets of hydrogen flashes of 1.35~$M_\odot$ WDs 
for three different mass-accretion rates, 
$\dot M_{\rm acc}=1.0 \times 10^{-10}~M_\odot$~yr$^{-1}$,  
$1.0 \times 10^{-8}~M_\odot$~yr$^{-1}$, and 
$5.0 \times 10^{-8}~M_\odot$~yr$^{-1}$. 
The structures for the first two accretion rates are
at the stages when the nuclear burning luminosity rises to 
$5.4 \times 10^3~L_\odot$ while the last one is at the slightly later
stage when to $8.0 \times 10^5~L_\odot$. 

Figure \ref{m135interior}(a) shows distributions of the pressure 
(black lines) and temperature (red lines).
The pressure hardly depends on the mass-accretion rate. 
The temperature is determined from the energy balance between  
gravitational energy release and neutrino loss. As shown in 
the figure, the temperature is almost isothermal throughout 
the WD interior owing to efficient conductive energy transport. 
The gravitational energy release rate is larger
for higher mass-accretion rates, so the resultant temperature
is higher in the larger mass-accretion models. 

Figure \ref{m135interior}(b) shows the degeneracy, $\psi$, i.e., the
chemical potential of an electron in units of $k T$, where $k$ is
the Boltzmann constant.  The degeneracy $\psi$ takes a large positive
value for strongly degenerated matter, and $\psi \lesssim 0$
for non-degenerated matter.  The WD interior is strongly degenerated 
for the lowest mass-accretion rate  
($\dot M_{\rm acc}=1.0 \times 10^{-10}~M_\odot$~yr$^{-1}$)  
while moderately degenerated in the other two cases. 

In the very surface region, the degeneracy is much weaker or almost 
lifted as shown by the red lines in Figure \ref{m135interior}(c)(d)(e). 
The boundary of the accreted H-rich matter and the 
CO WD is indicated by the black dot on the radius distribution line. 

Figure \ref{m135interior}(d) shows the interior structures (dotted lines)
at a slightly later stage when the nuclear burning increases to 
$L_{\rm nuc}=7.8 \times 10^5~L_\odot$. 
Nuclear energy (heat) flows toward both sides, not only outward but
also inward.  As a result, the degeneracy and density decrease
and the H-rich envelope starts to expand from the place indicated by
the black dot.

\subsection{Summary: Difference Between Classical Novae and Recurrent Novae}

A recurrent nova is a nova with multiple recorded outbursts.
Its recurrence time is less than a century. 
From figure \ref{dmdtmig}, a short recurrence time corresponds to the mass 
accretion rate of $\dot M_{\rm acc} \gtrsim 10^{-8}~M_\odot$~yr$^{-1}$ and 
very massive WDs ($\gtrsim 1.3~M_\odot$).  
On the other hand,  many classical novae correspond to relatively 
less massive WDs with lower mass-accretion rates. 

Figures \ref{dmdtmig}, \ref{Twd}, \ref{WDradius}, and \ref{m135interior}  
tell us that even when the WD masses are the same, 
the difference in the mass-accretion rate causes a substantial 
difference in the WD internal structure. 
In recurrent novae, the entire WD is warmed by the large gravitational
energy release rate owing to the higher mass-accretion rate.
Then the central temperature is relatively high, 
the WD radius is slightly larger (expanded), the ignition mass is smaller, 
and the accreted H-rich matter is non-degenerate. 
Then, the resultant shell flashes are relatively weak. 

On the other hand, in classical novae with lower mass-accretion rates, 
the WD is relatively cooler and has a slightly smaller radius $R_{\rm b}$. 
The ignition mass is large, and the accreted matter is strongly degenerated.  
Such a WD will experience a stronger thermonuclear runaway. 
Because the envelope mass at ignition is larger, we expect a 
brighter optical peak and massive ejecta comparing with those in
a recurrent nova having the same WD mass as we see in the next section.

\section{Nova Light Curves Based on the Steady-State Approximation}
\label{section_steadystate}

A nova outburst evolution can be followed by a sequence consisting
of steady-state envelope solutions with wind mass-loss
and without wind mass-loss (static envelope solutions after the winds stop). 
The wind is accelerated deep interior around the Fe peak in the opacity 
($T \sim 2 \times 10^5$ K) \citep{kat94h}. 
We calculated optical light curves of free-free emission 
that comes from optically thin ejecta near outside the photosphere,
which can be approximated as $F_\nu \propto 
\dot M_{\rm wind}^2 / (v_{\rm ph}^2 R_{\rm ph})$ \citep{hac06k}. 
The calculated light curve decays as $F_\nu \propto t^{-1.75}$ 
until the brightness drops by about 5 mag from the peak, followed by 
a steeper decline of $F_\nu \propto t^{-3.5}$ \citep{hac06k}. 
This decline trend is independent of the WD mass 
and chemical composition of the envelope.  Therefore, \citet{hac06k} 
called this trend ``the universal decline law'' of nova light curves.

The optical light curve decays faster in more massive WDs 
with smaller $X$ and larger $Z$.  
Such parameter dependences are already discussed in detail in 
\citet{kat94h}, \citet{hac06k}, \citet{hac10k}, \citet{hac15k},
\citet{hac16b}, \citet{hac18a}, \citet{hac18b}, and \citet{hac19b}, but the 
dependence on the WD radius ($R_{\rm b}$) has not been clarified.  
In section \ref{sec_interior}, we see that the WD radius depends on the 
mass accretion rate even if the WD masses are the same; The radius is 
smaller (larger) for a lower (higher) mass accretion rate. 
Here we discuss the dependence of the light curve on the WD radius
for a 1.35 $M_\odot$ WD.

\begin{figure*}
  \includegraphics[width=12cm]{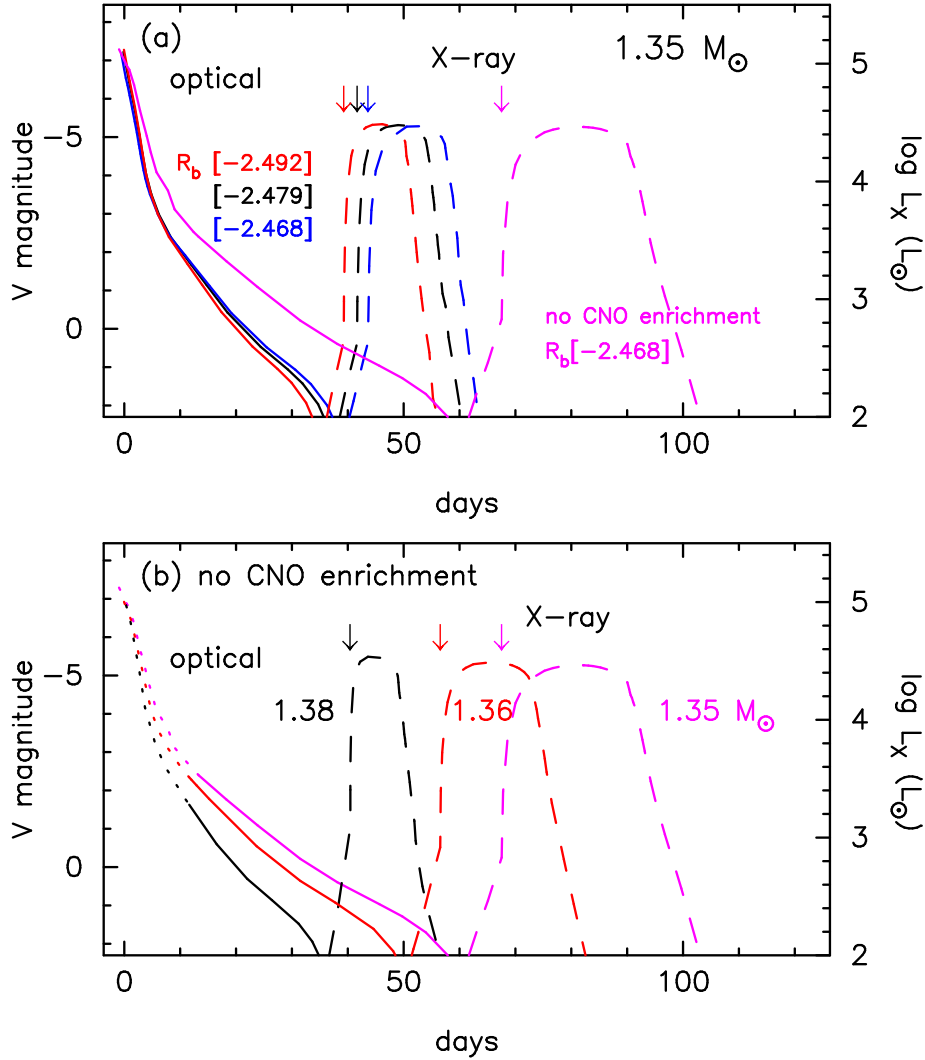}
\caption{
Theoretical light curve models of optical free-free emission (solid lines)
and X-ray blackbody emission (dashed lines) based on the steady-state
assumption. 
(a) Comparison of theoretical light curves for $1.35 ~M_\odot$ WDs 
with three different radii of $\log (R_{\rm b}/R_\odot)= -2.492$
(very cold WD: red), $-2.479$ (black), and $-2.468$ (blue).
The chemical composition of the envelope is assumed to be 
Ne nova 2 ($X=0.55,~Y=0.3,~X_{\rm CNO}=0.10,~X_{\rm Ne}=0.03$, $Z=0.02$
by mass fraction). 
For comparison we added the magenta lines of $X=0.55$, $Y=0.43$
and $Z=0.02$ with a radius of $\log (R_{\rm b}/R_\odot)= -2.468$ 
that may represent a case of recurrent nova like RS Oph.	
The downward arrow indicates the epoch when the optically thick 
wind stops in each model.  
(b) Comparison of the light curves for different WD masses but with the same 
recurrent nova composition ($X=0.55,~Y=0.43, ~Z=0.02$). In the dotted part of 
the optical light curve, the envelope mass exceeds the ignition mass 
for $\dot M_{\rm acc}=3 \times 10^{-8}~M_\odot$~yr$^{-1}$.
See the main text for more detail.  
\label{lightcurve.theory}}
\end{figure*}

\subsection{$1.35~M_\odot$ WD Models for Different Parameters}

Figure \ref{lightcurve.theory}(a) compares light curves 
of 1.35~$M_\odot$ WDs having various WD radii $R_{\rm b}$. 
The chemical composition is assumed to be
Ne nova 2 (25 \% mixing) from a grid chemical
composition model \citep{hac10k}, which seems to be close to 
the chemical composition obtained for V2491 Cyg by \citet{mun11},
i.e., $X = 0.573$, $Y = 0.287$, and $Z = 0.140$.  
The three X-ray light curves, which are closely located, 
are the models with the radius, from left to right, 
$\log R_{\rm b}/R_\odot= -2.492$ (red line), $-2.479$ (black),
and $-2.468$ (blue). 
These radii correspond to the one at ignition of 
a cold WD with $\dot M_{\rm acc}=5 \times 10^{-10}~M_\odot$~yr$^{-1}$, 
an expanded radius after the ignition, and the radius 
at ignition for $\dot M_{\rm acc}=5 \times 10^{-8}~M_\odot$~yr$^{-1}$,
respectively.

The three optical light curves (come from free-free emission) 
are very similar except in the later phase ($t \gtrsim 30$ d).
The free-free light curve is mainly determined by the wind mass-loss rate
$\dot M_{\rm wind}$. 
The wind is accelerated in the Fe peak region of OPAL opacity 
($T \sim 1.5 \times 10^5$ K), which is located far outside the 
nuclear burning region ($T \sim  10^8$ K). 
Thus, the wind mass-loss rate is not sensitive to the WD radius. 
In the later phase, the wind mass-loss rate decreases 
and becomes comparable to the mass decreasing rate due to nuclear burning. 
The nuclear burning rate is sensitive to the WD radius. 
For a smaller radius, the burning temperature is higher, so the
nuclear burning rate is larger. Thus, 
the light curve decays faster for a smaller WD radius. 

The X-ray flux quickly rises when the wind mass-loss stops 
(at $t \sim 40$ d) and decays after $\sim 10$ days. 
This X-ray turn on time ($t_{\rm on} \sim 40$ d) and the short SSS duration
($\Delta t_{\rm SSS} \sim 10$ d) are very consistent with the observed
X-ray turn-on time of V2491 Cyg and its duration.  In the next subsection,
we will make a composite light curve model of V2491 Cyg.

For comparison, we add a light curve model corresponding to a recurrent
nova with the same WD mass (1.35 $M_\odot$) but with a different 
chemical composition taken from the model in section \ref{sec_evolution}
(figure \ref{LnLg3mass}(b)), $X=0.55$, $Y=0.43$, and $Z=0.02$. 
This model shows much slower evolution because the wind acceleration is 
relatively weak due to the difference in the opacity. 
When the wind mass-loss stops, the envelope mass is 
larger by a factor of 1.6 compared with those in classical nova models. 
This makes the SSS duration longer than the classical nova models. 

Figure \ref{lightcurve.theory}(b) shows three WD models 
of different masses but with the same recurrent nova composition. 
The magenta line ($1.35 ~M_\odot$) model is the same as in the upper panel. 
In a more massive WD, the optical light curve decays faster and the SSS 
phase is shorter.
The 1.38 $M_\odot$ model shows an earlier X-ray turn-on time and 
shorter SSS duration, roughly consistent with the characteristic properties 
of V2491 Cyg. However, this model is based on the composition of 
$X=0.55$, $Y=0.43$, and $Z=0.02$, being inconsistent
with the estimates of heavy element enrichment \citep{mun11}. 
Thus, we reject the recurrent nova model for V2491 Cyg.

The X-ray turnoff time ($t \sim 90$ d) of the 1.35 $M_\odot$ model 
with the recurrent nova composition is roughly consistent 
with the recurrent nova RS Oph. 
However, the model X-ray flux rises much later 
and hence the model SSS phase is much shorter than the observed one. 
\citet{hac07kl} already reported this discrepancy in 
the steady-state model calculated for RS Oph. 
Here, we add another inconsistency between the envelope mass
and ignition mass. 
In the optical light curve shown in figure \ref{lightcurve.theory}(b), 
the envelope mass is $M_{\rm env}= 5 \times 10^{-6}~M_\odot$  
at the peak of the dotted part of the 1.35 $M_\odot$ model. 
This is much larger than the expected ignition mass for a recurrent nova, 
$M_{\rm ig} = 7 \times 10^{-7}~M_\odot$ 
at $\dot M_{\rm acc}=5 \times 10^{-8}~M_\odot$~yr$^{-1}$ 
as shown in figure \ref{dmdtmig}. 
If we adopt the small ignition mass ($M_{\rm env}= M_{\rm ig}=
7 \times 10^{-7}~M_\odot$), the optical light curve would start 
from the top of the solid part rather than the dotted part,
which is much fainter as a recurrent nova. 

To summarize, the steady-state model reproduces a consistent light curve for 
V2491 Cyg as a classical nova, but do not for the recurrent nova RS Oph. 
In section \ref{sec_evolution}, we discuss the reason why the steady-state
approximation fails to reproduce recurrent novae.

\subsection{V2491 Cyg Model}   \label{sec_v2491cyg}

\begin{figure*}
 \begin{center}
 \includegraphics[width=12cm]{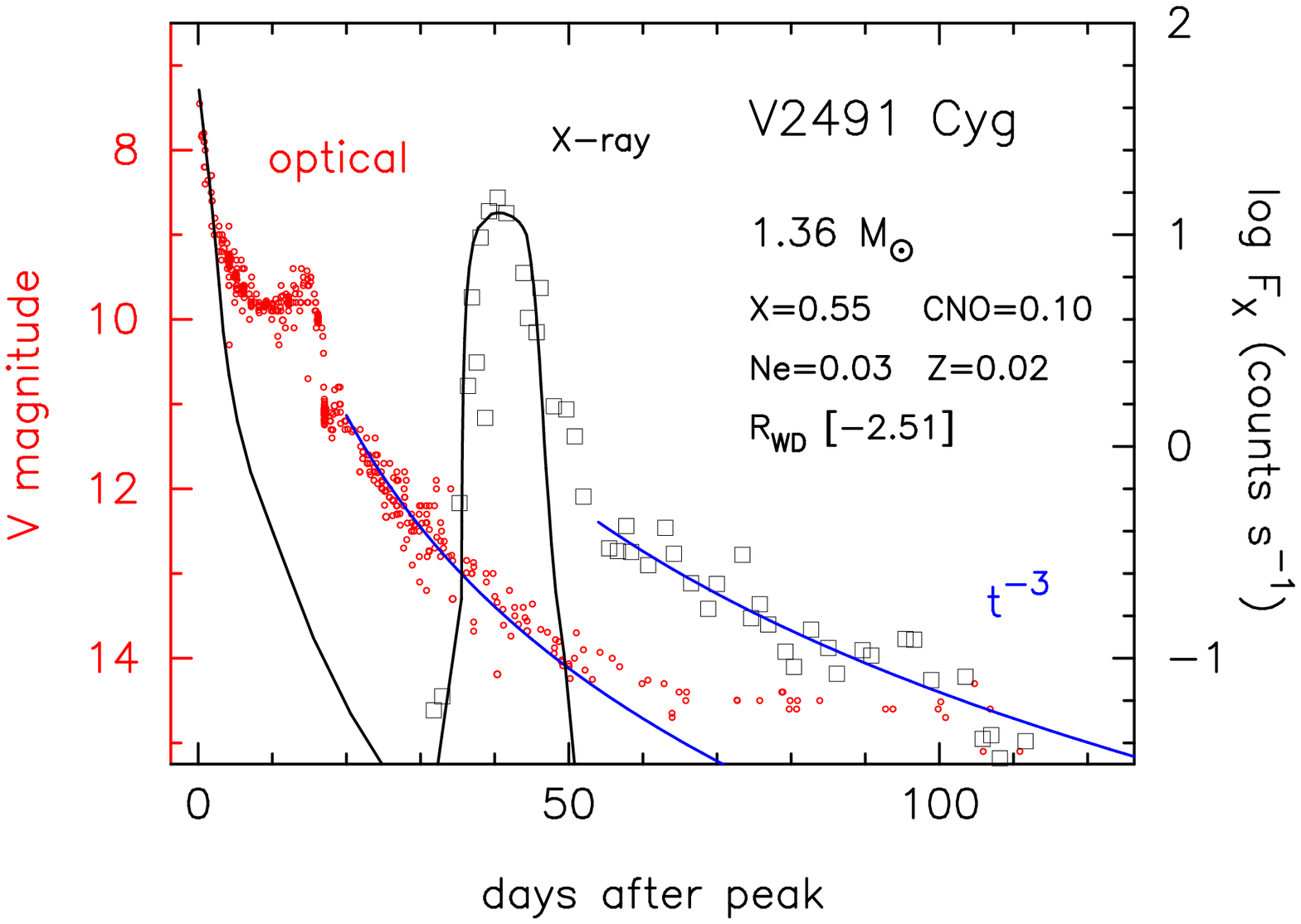}
\end{center}
\caption{
A model light curve fitting for V2491 Cyg. 
The thick black line indicates the optical (free-free emission) 
and X-ray (blackbody) light curves for a 1.36 $M_\odot$ WD with the 
radius of $\log (R_{\rm b}/R_\odot)=-2.51$.
The envelope chemical composition is Ne nova 2. 
The thin blue lines indicate the $F_\nu \propto t^{-3}$ law 
that represents free-free emission from freely expanding plasma 
with a constant mass, ejected on Day 0.  
\label{v2491cyg.f}}
\end{figure*}

We have calculated a light curve model for V2491 Cyg, assuming that
the WD mass is 1.36 $M_\odot$, its radius is 
$\log (R_{\rm b}/R_\odot)=-2.51$, and the envelope chemical composition 
is Ne nova 2. The radius $R_{\rm b}$ is indicated by the filled blue 
triangle at $\dot M_{\rm acc}=1 \times 10^{-11}~M_\odot$~yr$^{-1}$
in figure \ref{WDradius}. 
 This mass-accretion rate is obtained assuming that
the envelope mass at the optical peak, which is calculated later
in the light curve fitting for V2491 Cyg, is the same as that of
the ignition mass \citep{hac20}.
Figure \ref{v2491cyg.f} shows our light curve model. 
The optical free-free emission light curve 
is fitted to the observational 
data assuming $(m-M)_V=16.65$ \citep{hac21k}. 
The X-ray luminosity calculated with the blackbody assumption 
is arbitrarily shifted in the vertical direction to fit
the observed X-ray count rates. 

The theoretical light curve decays as $F_\nu \propto t^{-1.75}$, 
and fits the observational data only in the very early phase
down to $V \sim 9$. 
Such a rapid decay is due to the massive WD and heavy element enrichment 
as shown in the previous subsection. 
The calculated wind mass-loss rate is largest 
at the optical peak, $\dot M_{\rm wind}=3.6 \times 10^{-4}~M_\odot$~yr$^{-1}$, 
and quickly decreases with time until the wind stops at $t=35.7$ d. 
This quick decrease of the wind mass-loss rate is the reason for 
the steep decline of the optical light curve. 

Here we do not consider the second peak at Day $\sim 15$, 
which can be explained by some additional mechanism like 
the magnetic energy release \citep{hac09}. 
After the second peak, the theoretical curve 
is much fainter than the observed one. 
From our experience of a number of light curve fittings, 
nova brightnesses sometimes show much excess than the theoretical one
of $F_\nu \propto t^{-1.75}$. 
One of the reasons is shock heating. 
\citet{hac18b} examined light curves of fast novae and recurrent novae and 
showed that shock heating makes the light curve decay slower
than $F_\nu \propto t^{-1.75}$.  
For example, in the symbiotic classical nova V407 Cyg, 
the V magnitude decays as $F_V \propto t^{-1}$ 
until the shock breakout. 
In RS Oph, the Solar Mass Ejection Imager (SMEI) magnitude decays
as $F_{\rm SMEI} \propto t^{-1}$, while the V magnitude decays as
$F_V \propto t^{-1.55}$. 
The other effect is nebular line emission. 
In many classical novae observed data follow the theoretical line 
but becomes much brighter in $V$ magnitude 
when they enter the nebular phase.  
It is because the ejecta are ionized by high energy radiation
from the hot WD and many emission lines contribute to the $V$ magnitude. 
Typically the light curve roughly decays as $F_\nu \propto t^{-3}$
in the nebular phase (e.g. V1974 Cyg in \citet{woo97};
LV Vul in figure 24, V1500 Cyg in figure 45, both of \citet{hac19b}).  

In the case of V2491 Cyg, it is unlikely that the companion is a red giant,
and therefore we do not expect cool red-giant winds, as discussed
in section \ref{sec_rsophv2491}.
Instead, we expect that massive ejecta could contribute to the 
optical $V$ magnitude. In the model in figure \ref{v2491cyg.f}, 
the envelope mass at the optical peak ($t=0$ d) 
is $M_{\rm env}=2.4 \times 10^{-6}~M_\odot$, 
which is quickly reduced by the wind mass-loss to 
$M_{\rm env}=3.3 \times 10^{-7}~M_\odot$ 
until $t=10.3$ d, the beginning of the second peak, and 
to $M_{\rm env}=7.4 \times 10^{-8}~M_\odot$ 
at $t=35.7$ d when the optically thick winds stopped.  
Thus, 90\% of the envelope mass is ejected until the second peak. 
If we approximate the ejecta mass to be constant after the second peak, 
the $V$ brightness decays approximately along the line of 
$F_V \propto t^{-3}$ in figure \ref{v2491cyg.f}.  
The decline rate of $t^{-3}$ represents emission from optically thin plasma, 
the mass of which is constant in time, but expands freely
(see, e.g., equation (1) of \citet{woo97}).  
The model line of $F_V \propto t^{-3}$
shows a good agreement with the observed $V$ magnitudes.
The photospheric temperature of the WD rises from $\log T$ (K)=4.00
at $t=0$ d, to 4.95 at $t=10.3$ d, and 5.54 at $t=35.7$ d.

The supersoft X-ray flux (0.3 - 1.0 keV) is calculated
from the blackbody emission at the WD photosphere. 
The predicted X-ray right curve quickly rises just after 
the wind stops at Day 35.7. After staying on the maximum 
level for $\sim 10$ days, 
it decays quickly at Day $\sim 50$
when nuclear burning is extinguished. 
On the other hand, the observed flux decays slowly after Day 50. 
To explain this slow decay we added another line of $t^{-3}$. 
The model line of $F_X \propto t^{-3}$ shows a good fit with the X-ray
count rates after the SSS phase.  
The ejecta is heated by the WD and emit X-rays by thermal Bremsstrahlung
process (free-free emission).  This is essentially the same as the process
in the optical $V$ band.

The envelope mass at the optical peak 
$M_{\rm env}=2.1 \times 10^{-6}~M_\odot$ is consistent with 
the ignition mass at the mass accretion rate of
$\log \dot M_{\rm acc} (M_\odot$~yr$^{-1})=-11$ for 1.36 $M_\odot$ 
($M_{\rm ig}=2.0 \times 10^{-6}~M_\odot$) in figure \ref{dmdtmig}.
This ignition mass is marked with the filled triangle
in figure \ref{dmdtmig}. 
This low mass-accretion rate is consistent with a theory of Maximum
Magnitude vs. Rate of Decline (MMRD) \citep{hac20}
as mentioned in section \ref{sec_rsophv2491}
and our assumption of a cooler WD having a smaller $R_{\rm b}$. 
In this way, our classical nova model consistently explains 
the optical and X-ray light curves of V2491 Cyg.

\section{Comparison with Evolution Models}
\label{sec_evolution}

\begin{figure*}
 \begin{center}
 \includegraphics[width=14cm]{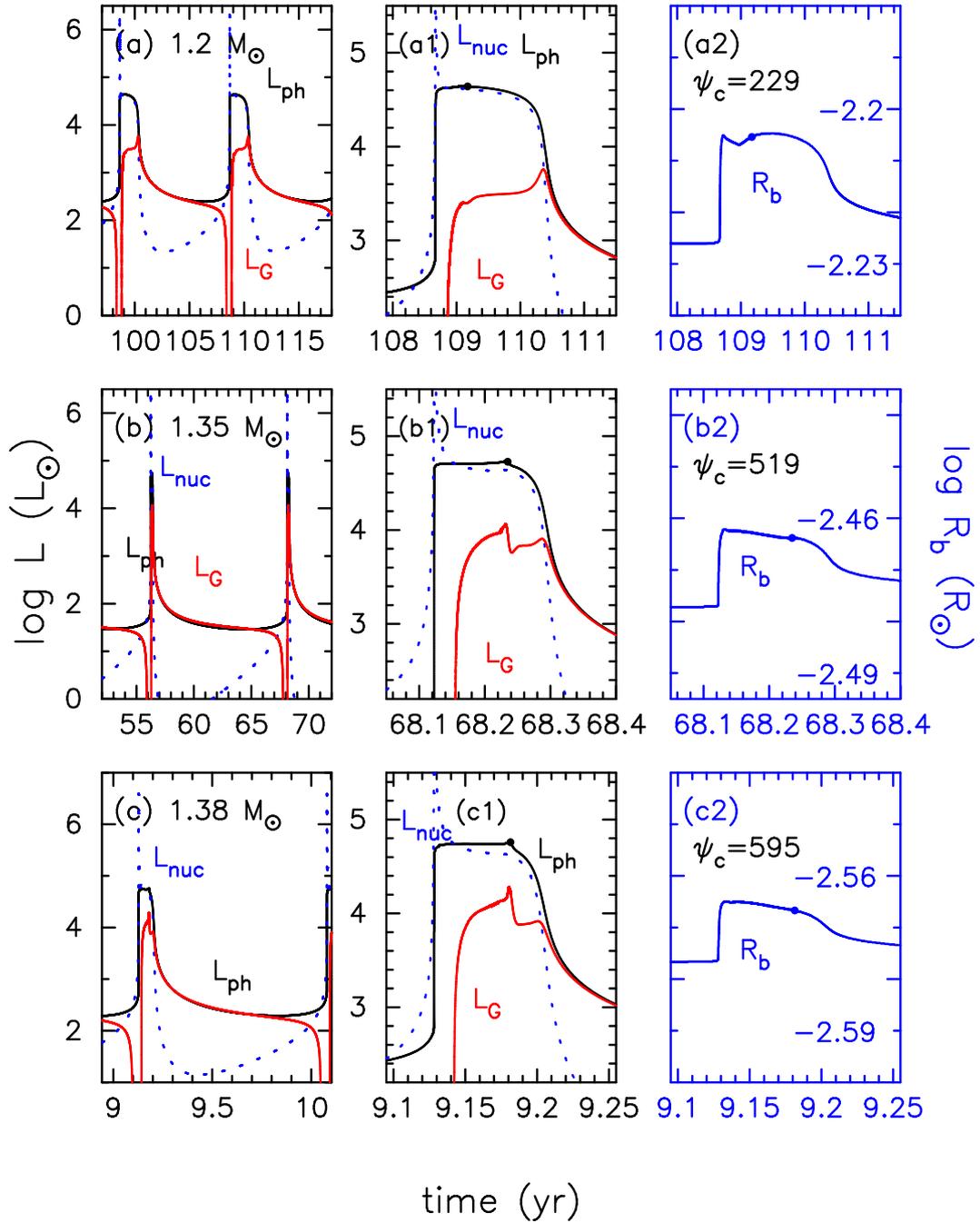}
 \end{center}
\caption{
Evolution models of (a) 1.2 $M_\odot$ WD with $\dot M_{\rm acc} 
= 2 \times 10^{-7}~M_\odot$~yr$^{-1}$, 
(b) 1.35 $M_\odot$ WD with $\dot M_{\rm acc} 
= 5 \times 10^{-8}~M_\odot$~yr$^{-1}$, 
and (c) 1.38 $M_\odot$ WD with $\dot M_{\rm acc} 
= 1.6 \times 10^{-7}~M_\odot$~yr$^{-1}$.
The left column shows the temporal changes of the photospheric luminosity
$L_{\rm ph}$ (black line), gravitational energy release rate 
$L_{\rm G}$ (red line), and nuclear burning rate $L_{\rm nuc}$
(dotted blue line) of the last two cycles of our calculation.    
The middle column shows close-up views of them. 
The black dot represents the stage when the wind mass-loss stops,
i.e., the beginning of the SSS phase. 
The right column shows the radius of the bottom of nuclear burning region,  
the scale of which is indicated inside of the figure. 
The degeneracy at the center of the WD, $\psi_{\rm c}$, is 
also indicated.  
\label{LnLg3mass}}
\end{figure*}

In the previous section, we see that the steady-state model explains 
the optical and X-ray light curves of V2491 Cyg.  
A good way to evaluate the steady-state approximation 
is to compare their internal structure with evolution models.
However, no internal structures have been published in evolution code
calculations for classical novae after the opacity tables are
revised in 1990's (see \citet{kat17palermo} for more details). 
Therefore, we concentrate on recurrent novae and compare 
our Henyey-type code solutions with steady-state solutions 
to see the reason why the light curve of RS Oph is different from
that of V2491 Cyg.

\subsection{Three Recurrent Nova Models}\label{sec_3RN}

Figure \ref{LnLg3mass} shows our time-dependent models of a 1.2 $M_\odot$ WD
with $\dot M_{\rm acc} =2 \times 10^{-7}~M_\odot$~yr$^{-1}$ 
($P_{\rm rec}=9.9$~yr), 
1.35 $M_\odot$ WD with $\dot M_{\rm acc}=5 \times 10^{-8}~M_\odot$~yr$^{-1}$ 
($P_{\rm rec}=12$~yr), 
and 1.38 $M_\odot$ WD with 
$\dot M_{\rm acc} = 1.6 \times 10^{-7}~M_\odot$~yr$^{-1}$ 
($P_{\rm rec}=0.95$~yr).  
The calculation method and the results for the 
1.2  $M_\odot$ and 1.38  $M_\odot$ models are
already published in \citet{kat17sha}. 
The left column in figure \ref{LnLg3mass} shows the last cycles of each
sequence, and the middle column shows the close-up view of the last flash, 
i.e., the change of the photospheric luminosity $L_{\rm ph}$, 
total nuclear burning rate $L_{\rm nuc}$, and gravitational energy
release rate $L_{\rm G}$.  The exact definition of $L_{\rm nuc}$
and $L_{\rm G}$ are given in \citet{hac16sk}. 
During the early outburst, a large amount of nuclear energy
$L_{\rm nuc}$ $(> 10^6~L_\odot)$ is generated, much larger than
the Eddington luminosity $L_{\rm Edd}=4 \pi c GM / \kappa$.
Most of the energy is absorbed but a part of it flows outward
to the photosphere. 
As a result, the photospheric luminosity $L_{\rm ph}$ 
does not exceed but is almost equal to the Eddington luminosity. 
This absorbed energy flux, expressed as a negative value of $L_{\rm G}$, 
is released shortly later, and $L_{\rm G}$ turns to be positive. 

The rightmost panels show the change of the radius at the bottom of hydrogen 
nuclear burning region $R_{\rm b}$. 
The radius $R_{\rm b}$ quickly rises at the onset of thermonuclear runaway 
because the envelope is heated and expanded, resulting in the 
decrease of pressure at $R_{\rm b}$.  
The radius keeps almost constant with gradual decrease in the temperature,
followed by quick decrease when $L_{\rm nuc}$ drops quickly. 
We indicate the degeneracy at the center of the WD, $\psi_{\rm c}$, 
as a representative value of thermal condition for the WD interior as
shown in figure \ref{m135interior}(b). 
The logarithmic radius increase $\Delta \log R_{\rm b}$ is larger 
for a smaller central degeneracy $\psi_{\rm c}$,
because weakly degenerated matter is 
easily expanded than in the tightly degenerated case 
(A shell flash occurs even if the envelope is not degenerated. 
See discussion in section \ref{shellflash}).

\subsection{Envelope Structure} \label{sec_struc}

\begin{figure*}
 \begin{center}
\includegraphics[width=17cm]{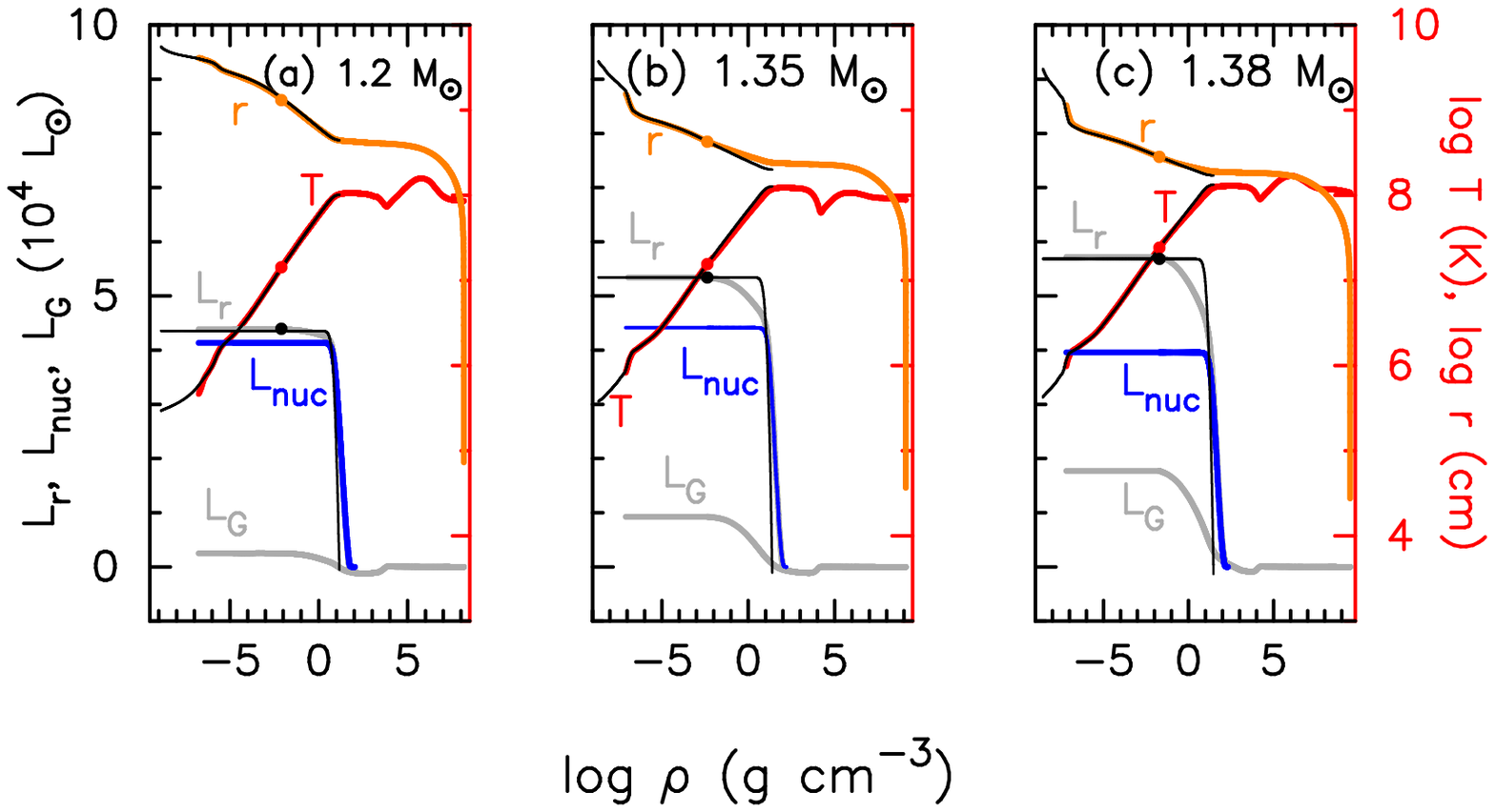}
 \end{center}
\caption{
Internal structures of the WDs obtained with our Henyey code, 
(a) 1.2 $M_\odot$ WD, (b) 1.35 $M_\odot$ WD, and 
(c) 1.38 $M_\odot$ WD models
at the epoch when the wind mass loss just stops, which are
indicated by the dot in figure \ref{LnLg3mass}. 
The radius $r$, temperature $T$, radiative flux $L_{\rm r}$, integrated
nuclear burning rate $L_{\rm nuc}$, and integrated gravitational energy
release rate $L_{\rm G}$ are plotted against the density. 
We omit the inner part of the $L_{\rm nuc}$ line, where it becomes zero, 
to clarify the negative value of $L_{\rm G}$.  
The thin black lines represent those of the hydrostatic solution
of steady-state approximation. 
The matching point with the Henyey code solution is indicated by the dot.
See the main text for more details.  
\label{ulg.3mass}}
\end{figure*}

Figure \ref{ulg.3mass} shows the distribution of $r$, $T$, $L_{\rm r}$, 
$L_{\rm nuc}$, and $L_{\rm G}$ against the density at the epoch 
when the wind mass loss stops as indicated by the dot
in figure \ref{LnLg3mass}. 
The rightmost point is the center of the WD. 
Hydrogen nuclear burning occurs at $\log \rho$ (g~cm$^{-3}$) $=1 - 2$,
where the nuclear luminosity $L_{\rm nuc}$ quickly increases 
outward (leftward). Convection does not occur in this stage. 
Heat flows not only outward but also inward. As a result, 
the temperature distribution has a temporal flat peak 
at $\log \rho$ (g~cm$^{-3}$) $=1.5 - 3$, that disappears 
until the end of the shell flash. 
In the inner region to the temporal peak, the temperature profile hardly 
changes with time, regardless the phase of the shell flash. 
The temperature slightly decreases at the center of the WD 
because of neutrino energy loss.

The gravitational energy release represents 
non-homologous (local) response of the envelope to heat generation/transport. 
It takes negative values when the envelope absorbs 
energy to expand while positive when loses energy to shrink. 
The integrated gravitational energy release rate, $L_{\rm G}$, 
increases outward in the region $\log \rho$ (g~cm$^{-3}$) $=-2$ - $+2$ 
and becomes constant outside this density region. 
This means that the inner part of the envelope is 
locally shrinking to release gravitational energy.   
This $L_{\rm G}$ term can be calculated only in the evolution code and is 
not included in hydrostatic steady-state solutions. 
Thus, the amount of $L_{\rm G}$ is a measure 
of the deviation from hydrostatic steady-state structure.

The radiative luminosity is the summation of the nuclear luminosity and
gravitational energy release rate, $L_{\rm r}=L_{\rm nuc}+L_{\rm G}$,
because neutrino loss is very small.  For the three cases
in figure \ref{ulg.3mass}, $L_{\rm G}$ is larger in more massive WDs and  
as large as about a half of $L_{\rm nuc}$ in the 1.38 $M_\odot$ WD. 
Note, however, that $L_{\rm G}$ takes the maximum value at this epoch 
in 1.35 $M_\odot$ and 1.38 $M_\odot$ WDs
(see figure \ref{LnLg3mass}) and decreases after that.

\subsection{Comparison with Hydrostatic Solutions of Steady-state Approximation}

Figure \ref{ulg.3mass} compares the internal structures of 
evolution models calculated with a Henyey code (thick solid lines) 
with hydrostatic steady-state solutions (thin solid black lines). 
The two solutions are fitted at the matching point (the filled circles)  
to have the same temperature, radius and luminosity $L_{\rm r}$ 
(See \citet{kat17sha} for more details). 

In the 1.2 $M_\odot$ WD, the temperature $T$ and radius $r$ of
these two different solutions are very close to each other not only
in the outside region to the matching point but also inside region
down to the place where $L_{\rm nuc}=0$. 
In the 1.35 $M_\odot$ and 1.38 $M_\odot$ WDs,
the hydrostatic steady-state solution fits well with the evolution
solution in the $r$, $T$, and $L_{\rm r}$ distributions in the outside
region to the fitting point. 
Although the hydrostatic steady-state solutions do not include $L_{\rm G}$, 
we obtain a rather good agreement with the $L_{\rm r}=L_{\rm nuc}+L_{\rm G}$ 
evolution solutions.  In other words, the static steady-state solution is
close to those of evolution solution even when $L_{\rm G}$
substantially contributes to the total luminosity
as far as $L_{\rm r}$ is constant outside the fitting point.

In the time-dependent solution, $L_{\rm G}$ decreases inward  
differently from $L_{\rm nuc}$.
This causes the large deviation in the $L_{\rm r}$ distribution, 
and also in the envelope mass distribution. 
Thus, the evolution time is different between evolution calculation 
and steady-state wind and hydrostatic sequence. 
We may say that steady-state wind/hydrostatic treatment is a good
approximation when $L_{\rm G}$ is small compared with $L_r$
because the envelope structure is close to that of evolution solution,
but not always good when $L_{\rm G}$ is comparable to $L_r$. 
We will discuss this point in section \ref{sec_steadystate}.

\section{DISCUSSION}\label{sec_discussion}

\subsection{Steady-State Approximation} \label{sec_steadystate}

We calculated the light curve model with steady-state approximation 
in section \ref{sec_v2491cyg}.  This model explains the observed
light curves of the classical nova V2491 Cyg.  
On the other hand, the model light curves for the recurrent nova composition 
(magenta line in figure \ref{lightcurve.theory})
does not fit to the X-ray light curve of RS Oph. 
These two novae host similar WD masses ($\sim 1.35-1.36~M_\odot$), 
but with very different mass-accretion rates. 
Here, we discuss the reason why the steady-state assumption
is good for classical novae (with lower mass-accretion rates), 
but not for recurrent novae (with higher mass-accretion rates), 
even with similar WD masses.

Kato's (1983, 1985) optically thick wind theory is to follow
a nova outburst with a sequence of steady-state wind and static
solutions (see also \citet{kat94h} for OPAL opacity).
The time between two consecutive solutions is calculated
from the difference in the envelope mass and its decreasing rate
due both to wind mass-loss and nuclear burning.  After the wind stops,
the envelope mass decreases only with nuclear burning.

In the SSS phase, we calculate the envelope 
mass-decreasing rate based on the nuclear luminosity,  
$L_{\rm nuc} =L_{\rm ph}$.  We would overestimate the mass-deceasing rate
when $L_{\rm G}$ substantially contributes to the total luminosity, i.e.,
$L_{\rm nuc} < L_{\rm ph}$.
Thus, the SSS phase could be shorter than the evolution calculation,
as expected from figure \ref{LnLg3mass}(a)(b)(c).

When strong optically-thick winds occur in the early phase of a nova
outburst, the timescale is determined mainly by the wind mass-loss rate
because the total mass decreasing rate is determined mainly
by the wind mass-loss rate not by the nuclear burning rate.
Thus, the effect of $L_{\rm G}$ on the timescale is negligible even if
$L_{\rm G}$ substantially contributes to the total luminosity.

In high mass-accretion rates, the contribution of $L_{\rm G}$ is not
negligible as already shown in figure \ref{LnLg3mass}(a)(b)(c).
If its contribution is very small or negligible in low mass-accretion rates,
we may conclude that the steady-state approximation
is good for classical novae.  In order to check this question,
we searched literature for hydrogen shell flashes having the description
on the contribution of $L_{\rm G}$ to the interior structures. 
Many shell flash calculations have been published so far,
but most of them are lack of information on $L_{\rm G}$. 
Among them, \citet{ibe82} described energy balance in the WD core
between the energy gain owing to gravitational heating and nuclear burning
and energy loss by neutrino. 
In the case of $1.01 ~M_\odot$ WD with 
$\dot M_{\rm acc}= 1.5 \times 10^{-8}~M_\odot$~yr$^{-1}$, 
the gravitational heating is as small as $L_{\rm G}/L_{\rm nuc} < 10^{-4}$. 
For the case of $1.5 \times 10^{-9}~M_\odot$~yr$^{-1}$, 
$L_{\rm G}$ is not plotted but the photospheric luminosity  
$L_{\rm ph}$ is equal to $L_{\rm nuc}$ within the line width in the figure. 
Thus $L_{\rm G}=L_{\rm ph}-L_{\rm nuc}$ should be very small. 
These WD mass and mass accretion rates are typical for classical novae, 
and we regard that the steady-state approximation is good.

The above calculation is done before the opacity tables are
recalculated in the beginning of 1990's.  With the OPAL opacity \citep{igl96}, 
strong winds are accelerated in the region of $T \sim 10^5$ K. 
This region is much outside of the nuclear burning region ($T \sim 10^8$ K) 
(see e.g., \citet{kat94h}).  
Thus, the opacity does not much affect the deep inside 
where $L_{\rm G}$ is mainly generated. 
We may conclude that $L_{\rm G}$ is also small in shell flash calculation 
even with the OPAL opacity.

To summarize, the gravitational energy release rate $L_{\rm G}$ 
is very small or negligible in low mass-accretion rates while this
effect on the timescale should not be neglected in the SSS phase of
high mass-accretion rates.  Thus, we may conclude that the steady-state
assumption is a good approximation for the classical nova V2491~Cyg.

\subsection{Shell Flashes in Non-degenerate Envelope} \label{shellflash}

A hydrogen shell flash could occur regardless whether the envelope matter 
is degenerate or not. 
Figure \ref{m135interior} shows that the envelope is non-degenerate 
for high mass-accretion rates. 
\citet{jos93} presented the shell flash calculation with two-zone
approximation in which the temperature in the quiescent phase is as high as 
$T = 3.4 \times 10^{7}$ K ($\log \rho$ (g~cm$^{-3})\sim 1.8$). 
In the model of a $1.2~M_\odot$ WD with
$\dot M_{\rm acc} =2 \times 10^{-7}~M_\odot$~yr$^{-1}$, 
the bottom of the H-rich envelope is non-degenerate 
throughout a full cycle of the shell flash.

\citet{sug78} discussed that the most essential factor of thermonuclear
runaway is the plane parallel configuration of the envelope structure.
When a shell flash begins, nuclear burning produces thermal energy, and 
both the temperature and pressure at the nuclear burning region increase. 
This causes a shell expansion. If the envelope structure is spherical, 
the expansion of the shell results in the pressure decrease.  Then,
the temperature also decreases. 
Thus, the nuclear burning is stabilized and runaway does not occur. 
In the plane parallel configuration, however,
the pressure at the bottom of the envelope is written as  
$P_{\rm b}= G M_{\rm WD} M_{\rm env} /(4 \pi R_{\rm b}^4)$, i.e.,  
the pressure  is determined only by the envelope mass 
$M_{\rm env}$ and independent of the temperature. 
When a shell flash occurs keeping the plane parallel structure, 
the bottom pressure hardly changes, so that the temperature would increase 
and accelerate nuclear reactions.  This causes the thermonuclear runaway. 
If the matter is strongly degenerate, the pressure is independent of the 
temperature.  The strong degeneracy enhances and accelerates the nuclear
burning, but not the necessary condition for shell flashes.
The most essential factor for the thermonuclear runaway
is the plane parallel configuration.

\subsection{WD Core Temperature}
\label{discussion_Tc}

In a long-term evolution of an accreting WD, its interior 
approaches a thermally relaxed state in which the energy gain 
owing to gravitational contraction of the core and time-averaged
nuclear burning balances with the energy loss by neutrino. 
\citet{ibe82} obtained such a core temperature of a WD with double shell
burning.  His accreting WD core is in a thermal balance in which
``the release of gravitational potential energy within the core is
balanced to go into neutrinos and into rising kinetic energy of electrons 
and nuclei in the core.''  
The equilibrium core temperature is obtained to be 
$\log T_{\rm c}$ (K)= 7.84 (1.01 $M_\odot$), 
7.90 (1.20 $M_\odot$), and 8.06 (1.40 $M_\odot$) for 
$\dot M_{\rm acc}=1.5 \times 10^{-8}~M_\odot$~yr$^{-1}$. 
We have calculated a similar case 
using the same method and numerical code in the present work and 
obtained the WD core temperature to be 
$\log T_{\rm c}$ (K)= 7.87 (1.1 $M_\odot$), 7.88 (1.20 $M_\odot$), 
and 7.96 (1.38 $M_\odot$) for $\dot M_{\rm acc}=2.0 \times 10^{-8}~M_\odot$~yr$^{-1}$. 
These temperatures are in reasonable agreement with Iben's results. 
As the burning temperature of helium is much higher than 
that of hydrogen, these core temperatures are higher than the values 
in figure \ref{Twd}. 

\citet{tow04} calculated the equilibrium core temperature of classical 
novae, as a function of the accretion rate and accreted mass 
for a specified WD mass.  For example, starting from a 0.6 $M_\odot$ WD
with $\dot M_{\rm acc}=4\times 10^{-11}~M_\odot$~yr$^{-1}$, they obtained 
$T_{\rm c} = 1.5 \times 10^7$ K
for the accreted mass $1 \times 10^{-4}~M_\odot$.   
This value is much lower than our results of 
$T_{\rm c}=2.36 \times 10^7$ K and $M_{\rm ig}=2.57\times 10^{-4}~M_\odot$ 
for the same WD mass and mass accretion rate. 
We do not clarify the reason, 
but this difference probably comes from the approximations  
adopted in their calculation.

\citet{pie00} calculated hydrogen shell flashes for a 0.5168 $M_\odot$ WD  
with $\dot M_{\rm acc}=1.0 \times 10^{-8}~M_\odot$~yr$^{-1}$ starting from 
a cold WD of $\log T_{\rm c}$ (K)=6.6017 which is raised to 
6.9355 after only successive 60 hydrogen shell flashes  
(the WD mass increases to 0.5235 $M_\odot$).

\citet{epe07} also showed that the initially hot (cold)
WD of $T_{\rm c}=5 \times 10^7$ K ($ 0.5 \times 10^7$ K) 
cools down (becomes hot), after the 0.65 $M_\odot$ WD experienced 
$\sim 3000$ shell flashes ($10^8$~yr) 
with $\dot M_{\rm acc}= 1\times 10^{-9}M_\odot$~yr$^{-1}$. 
In the both cases of hot and cold WDs, $T_{\rm c}$ seems to approach
a single value of $\sim 1.5 \times 10^7$ K, 
which is much lower than our results of $3.5 \times 10^7$ K,
probably because they did not include heat balance of the core 
between gravitational energy release and neutrino loss, and because of 
some difficulties in their numerical code 
(see \citet{kat17palermo} for more details). 

In these evolution calculations (except for \citet{tow04}),
the flash properties such as the recurrence period, 
flash strength, and amount of mass-loss drastically changed 
in successive shell flashes (e.g., \citet{epe07} and \citet{hil16}). 
To avoid such an initial-model dependency,
we have started a calculation from an initial model close to a thermal
equilibrium as done in \citet{kat17shb}. 

Recurrent novae are one of the candidate of Type Ia supernova progenitors 
\citep{hkn96,hknu99,hkn99,kat12Review}. 
The mass-accreting WD grows its mass toward the Chandrasekhar mass limit. 
In such WDs the central temperature gradually increases keeping the 
equilibrium temperature as high as $> 10^8$ K 
for $M_{\rm WD} > 1.3 ~M_\odot$ (e.g. \citet{wan17,wu17}).  
In such a binary evolution, the mass retention efficiency is
an essentially important factor. 
If we adopt a very low initial temperature ($\sim 10^7$~K)
such as those assumed in 
\citet{pri95}, \citet{yar05}, \citet{sta12}, \citet{wol13}, and \citet{che19},
some of which are depicted in figure \ref{Twd}, the resultant nova properties  
and mass retention efficiency would be very different. 
For a cooler WD the flash is stronger and then the mass retention 
efficiency is estimated to be smaller than in a realistic (equilibrium)
case for a long-term binary evolution. 
It affects the long-term binary-evolution scenarios toward type Ia supernovae.


\section{Conclusions}\label{section_conclusion}

Our main results are summarized as follows.

\begin{enumerate}

\item
In a long-term cataclysmic binary evolution,
the WD interior is adjusted to the mass accretion, 
and the central temperature is not a free parameter. 
We calculated such a WD interior and obtained the ignition mass, WD radius, 
and WD central temperature for various WD masses and mass-accretion rates. 

\item
In classical novae, which correspond to the case of low mass-accretion rates, 
its WD interior is relatively cool and the ignition mass is larger. 
The envelope is strongly degenerated, 
thus, a stronger shell flash is expected. 
In recurrent novae, which correspond to massive WDs with 
high mass-accretion rates, its envelope is less or non-degenerated.
Thus, the shell flash is relatively weaker. 
Even if the WD masses are the same, their outburst properties are 
very different between high (recurrent novae) and low 
(classical novae) accretion rates.

\item 
The steady-state approximation is adequate to follow classical nova
outbursts but not for recurrent novae, because the gravitational
energy release rate $L_{\rm G}$ substantially contributes to 
change the envelope structures in recurrent novae.

\item
The characteristic properties of visual and X-ray light curves of V2491 Cyg
are explained with a sequence of steady-state wind/static solutions. 
We present a model of a cold 1.36 M$_\odot$ WD with the envelope chemical 
composition of Ne nova 2. The optical light curve is explained by
free-free emission from the plasma outside the photosphere in the early phase. 
The X-ray light curve is reproduced by blackbody fluxes from 
the WD photosphere, followed by emission from expanding hot ejecta. 

\end{enumerate}


\begin{ack}
 We thank the anonymous referee for useful comments that
 improved the manuscript.
\end{ack}



%


\begin{thebibliography}{}


\bibitem[Brandi et al. (2009)]{bra09}
Brandi, E., Quiroga, C., Miko{\l}ajewska, J., Ferrer, O. E.,
\& Garc{\'i}a, L. G. 2009, \aap, 497, 815


\bibitem[Chen et al. (2019)]{che19}
Chen, H., et al. 2019, \mnras, 490, 1678

\bibitem[Denissenkov et al. (2013)]{den13} Denissenkov, P. A., Herwig, F., 
Bildsten, L., \& Paxton, B. 2013, \apj, 762, 8

\bibitem[Epelstain et al. (2007)]{epe07} 
Epelstain, N.,  Yaron, O., Kovetz, A. \& Prialnik, D. 2007, \mnras, 374, 1449

 
\bibitem[Jos\'e et al. (1993)]{jos93}
 Jos\'e, J., Hernanz, M., \& Isern, J. 1993, \aap, 269, 291 




\bibitem[Hachisu \& Kato (2006)]{hac06k}
Hachisu, I., \& Kato, M. 2006, \apjs, 167, 59



\bibitem[Hachisu \& Kato (2009)]{hac09}
Hachisu, I., \& Kato, M. 2009, \apj, 694, L103

\bibitem[Hachisu \& Kato (2010)]{hac10k}
Hachisu, I., \& Kato, M. 2010, \apj, 709, 680

\bibitem[Hachisu \& Kato (2015)]{hac15k}
Hachisu, I., \& Kato, M. 2015, \apj, 798, 76

\bibitem[Hachisu \& Kato (2016a)]{hac16a}
Hachisu, I., \& Kato, M. 2016a, \apjs, 223, 21


\bibitem[Hachisu \& Kato (2016b)]{hac16b}
Hachisu, I., \& Kato, M. 2016b, \apj, 816, 26

\bibitem[Hachisu \& Kato (2018a)]{hac18a}
Hachisu, I., \& Kato, M. 2018a, \apj, 858, 108

\bibitem[Hachisu \& Kato (2018b)]{hac18b}
Hachisu, I., \& Kato, M. 2018b, \apjs, 237, 4 

\bibitem[Hachisu \& Kato (2019a)]{hac19a}
Hachisu, I., \& Kato, M. 2019a, \apjs, 241, 4 

\bibitem[Hachisu \& Kato (2019b)]{hac19b}
Hachisu, I., \& Kato, M. 2019b, \apjs, 242, 18

\bibitem[Hachisu \& Kato (2021)]{hac21k}
Hachisu, I., \& Kato, M. 2021, \apjs, 253, 27

\bibitem[Hachisu et al. (2000)]{hkkm00}
Hachisu, I., Kato, M., Kato, T., \& Matsumoto, K. 2000,
\apjl, 528, L97 

\bibitem[Hachisu et al. (2006)]{hac06c}
Hachisu, I., Kato, M., Kiyota, S., et al. 2006, \apjl, 651, L141

\bibitem[Hachisu et al. (2007)]{hac07kl}
Hachisu, I., Kato, M., \& Luna, G. J. M. 2007, \apj, 659, L153

\bibitem[Hachisu et al. (1996)]{hkn96}
Hachisu, I., Kato, M., \& Nomoto, K. 1996, \apj, 470, L97 

\bibitem[Hachisu et al. (1999a)]{hkn99}
Hachisu, I., Kato, M., \& Nomoto, K. 1999a, \apj, 522, 487 



\bibitem[Hachisu et al. (1999b)]{hknu99}
Hachisu, I., Kato, M., Nomoto, K., \& Umeda, H. 1999b, \apj, 519, 314

\bibitem[Hachisu et al. (2016)]{hac16sk}
Hachisu, I., Saio, H., \& Kato, M.  2016, \apj, 824, 22 

\bibitem[Hachisu et al. (2020)]{hac20}
Hachisu, I., Saio, H., Kato, M., Henze, M., \& Shafter, A.W. 2020, \apj,
902, 91 

\bibitem[Henze et al. (2018)]{hen18} Henze, M. et al. 2018, \apj, 857, 68


\bibitem[Hillman et al. (2016)]{hil16} Hillman, Y., Prialnik, D., Kovetz, A.\& 
Shara, M. M. 2016, \apj, 819, 168



\bibitem[Iben (1982)]{ibe82} Iben, I., Jr. 1982, \apj, 259, 244 



\bibitem[Iglesias \& Rogers (1996)]{igl96}
Iglesias, C. A., \& Rogers, F. J. 1996, \apj, 464, 943

\bibitem[Iijima (2009)]{iijima09} Iijima, T. 2009, \aap, 505, 2871



\bibitem[Kato (1983)]{kat83}
Kato, M. 1983, \pasj, 35, 507

\bibitem[Kato (1985)]{kat85} Kato, M. 1985, \pasj, 37, 19


\bibitem[Kato (1999)]{kat99} Kato, M. 1999, \pasj, 51, 525





\bibitem[Kato \& Hachisu (1994)]{kat94h} Kato, M., \& Hachisu, I. 1994, \apj, 437, 802

\bibitem[Kato \& Hachisu (2009)]{kat09} Kato, M., \& Hachisu, I. 2009, \apj, 699, 1293

\bibitem[Kato \& Hachisu (2012)]{kat12Review}
 Kato, M., \& Hachisu, I. 2012, Bull.Astr.Soc,India, 40, 393

\bibitem[Kato \& Hachisu (2020)]{kat20}
Kato, M., \& Hachisu, I. 2020, PASJ, 72, 82

\bibitem[Kato et al. (2013)]{kat13hh}
Kato, M., Hachisu, I., \& Henze, M. 2013, \apj, 779, 19 

\bibitem[Kato et al. (2014)]{kat14shn}
Kato, M., Saio, H., Hachisu, I., \& Nomoto, K. 2014, \apj, 793, 136


\bibitem[Kato et al. (2017c)]{kat17palermo}
Kato, M., Hachisu, I., \& Saio, H.  2017c, 
in Proceedings of the Palermo Workshop 2017
on ``The Golden Age of Cataclysmic Variables and Related Objects - IV'',
ed. F. Giovannelli et al. (Trieste: SISSA PoS), 315, 56


\bibitem[Kato et al. (2017a)]{kat17sha}
Kato, M., Saio, H., \& Hachisu, I.  2017a, \apj, 838, 153
 
\bibitem[Kato et al. (2017b)]{kat17shb}
 Kato, M., Saio, H., \& Hachisu, I. 2017b, \apj, 844,143


\bibitem[Kato et al. (2020)]{kat20sh}  Kato, M., Saio, H., \& Hachisu, I., 2020, \apj, 892, 15


\bibitem[Kato et al. (2016)]{kat16xflash} Kato, M., et al. 2016, \apj, 830, 40



\bibitem[Knigge et al. (2011)]{kni11}
Knigge, C., Baraffe, I., \& Patterson, J. 2011, \apjs, 194, 28

 

\bibitem[Munari et al. (2011)]{mun11}
Munari, U., Siviero, A., \& Dallaporta, S. 2011, New Astronomy, 16, 209

\bibitem[Nakano et al. (2008)]{nak08}
Nakano, S., Beize, J., Jin, Z. -W., et al. 2008, IAU Circular, 8934, 1 




\bibitem[Ness et al. (2011)]{nes11}
Ness, J.-U., Osborne, J.P., Dobrotka, A., et al. \apj, 733, 70

\bibitem[Ness et al. (2007)]{nes07} Ness, J.-U., et al. 2007, \apj, 665, 1334




\bibitem[Page et al. (2010)]{pag10}
Page, K.L., Osborne, J.P., Evans, P.A., et al., 2010, \mnras, 401, 121


\bibitem[Pagnotta \& Schaefer (2014)]{pag14}
Pagnotta, A., \& Schaefer, B. E. 2014, \apj, 788, 164,

\bibitem[Piersanti et al. (2000)]{pie00}
Piersanti, L., Cassisi, S., Iben, I. Jr., \& Tornambe, A. 2000, \apj, 535, 932



\bibitem[Prialnik \& Kovetz (1995)]{pri95}
Prialnik, D., \& Kovetz, A., 1995, ApJ, 445, 789


\bibitem[Sala \& Hernanz (2005)]{sal05}
Sala, G., \& Hernanz, M. 2005, \aap, 439, 1061 



\bibitem[Starrfield et al. (2012)]{sta12}
Starrfield, S., et al.  2012, BASI, 40, 419

\bibitem[Sugimoto \& Fujimoto (1978)]{sug78}
Sugimoto, D., \& Fujimoto, M. Y. 1978, \pasj, 30,467


\bibitem[Sun et al. (2020)]{sun20}
Sun, B., et al. 2020, \mnras, 499, 3006

\bibitem[Takei et al. (2009)]{tak09}
Takei, D. et al. 2009, \apjl, 2009, 697, L54

\bibitem[Takei et al. (2011)]{tak11}
Takei, D. et al. 2011, PASJ, 63, S729

\bibitem[Tomov et al. (2008)]{tom08}
Tomov, T., Mikolajewski, M., Brozek, T., et al. 2008, ATel, 1485, 1 

\bibitem[Townsley \& Bildsten (2004)]{tow04}
Townsley, D. M., \& Bildsten, L. 2004, \apj, 600, 390


\bibitem[Warner (1995)]{war95}
Warner, B. 1995, Cataclysmic variable stars, Cambridge, 
Cambridge University Press

\bibitem[Wang et al (2017)]{wan17}
Wang, B., Podsiadlowski, P., \& Han, Z. 2017, \mnras, 472, 1593 

\bibitem[Wolf et al. (2013)]{wol13}
Wolf, W. M., Bildsten, L., Brooks, J., \& Paxton, B. 2013, \apj, 777, 136, 
(Erratum: 2014, ApJ, 782, 117)


\bibitem[Woodward et al. (1997)]{woo97}
Woodward, C. E., Gehrz, R.D., Jones, T.J., Lawrence, 
G.F., \& Skrutskie, M.F. 1997, \apj, 477, 817


\bibitem[Wu et al. (2017)]{wu17}
Wu, C., Wang, B., Liu, D. \& Han, Z. 2017, \aap, 604, A31

\bibitem[Yaron et al. (2005)]{yar05}
Yaron, O., Prialnik, D., Shara, M.M., \& Kovetz, A. 2005, \apj, 623, 398


\bibitem[Zemko et al. (2015)]{zem15} Zemko, P., Mukai, K., \& Orio, M. 
2015, \apj, 807, 61







\end{thebibliography}
\end{document}